\documentclass[preprint,12pt]{aastex}
\begin{document}
\newcommand{\Halpha}{H$\alpha$ }
\newcommand{\etal}{\mbox{et al.}}
\newcommand{\NHoo}{N_{\rm H}}
\newcommand{\NHxo}{N_{\rm x}}
\newcommand{\NHIo}{N_{\rm HI}}
\newcommand{\NHtwoone}{N$_{\rm 21cm}$ }
\newcommand{\NHII}{N_{\rm HII}}
\newcommand{\NHgx}{N_{\rm G}}
\newcommand{\acm}{ cm$^{-2}$ }
\newcommand{\as}{s$^{-1}$}
\newcommand{\Htwo}{H$_{2}$}
\newcommand{\LHB}{{\sc lhb}}
\newcommand{\ISM}{{\sc ism}}
\newcommand{\LISM}{{\sc lism}}
\newcommand{\MER}{{\sc mer}}
\newcommand{\ASCA}{{\it ASCA}}
\newcommand{\IRAS}{{\it IRAS}}
\newcommand{\ROSAT}{{\it ROSAT}}
\newcommand{\COBE}{{\it COBE}}
\newcommand{\DIRBE}{{\it DIRBE}}
\newcommand{\EUVE}{{\it EUVE}}
\newcommand{\pp}{\phn}
\newcommand{\ppp}{\phn\phn}
\newcommand{\pppp}{\phn\phn\phn}
\newcommand{\pq}{\,$\pm$\,}
\newcommand{\pd}{\phn\phn\phn\,---}
\newcommand{\plong}{\hspace{10pt}}
\newcommand{\gte}{$\infty$\phn}
\newcommand{\ZY}{0.3,1.0}
\newcommand{\ZZ}{0.5,1.0}
\newcommand{\Msun}{$M_{\odot}$}
\newcommand{\ayr}{y$^{-1}$}
\newcommand{\ARAA}[2]{ARA\&A, #1, #2}
\newcommand{\ApJ}[2]{ApJ, #1, #2}
\newcommand{\ApJL}[2]{ApJL, #1, #2}
\newcommand{\ApJSS}[2]{ApJS, #1, #2}
\newcommand{\AandA}[2]{A\&A, #1, #2}
\newcommand{\AandASS}[2]{A\&AS, #1, #2}
\newcommand{\AJ}[2]{AJ, #1, #2}
\newcommand{\BAAS}[2]{BAAS, #1, #2}
\newcommand{\ASP}[2]{ASP Conf.\ Ser., #1, #2}
\newcommand{\JCP}[2]{J.\ Comp.\ Phys., #1, #2}
\newcommand{\MNRAS}[2]{MNRAS, #1, #2}
\newcommand{\N}[2]{Nature, #1, #2}
\newcommand{\PASJ}[2]{PASJ, #1, #2}
\newcommand{\PASP}[2]{PASP, #1, #2}
\newcommand{\PRL}[2]{Phys.\ Rev.\ Lett., #1, #2}
\newcommand{\RPP}[2]{Rep.\ Prog.\ Phys., #1, #2}
\newcommand{\ZA}[2]{Z.\ Astrophs., #1, #2}
\newcommand{\tenup}[1]{\times 10^{#1}}
\newcommand{\pz}[0]{\phantom{0}}

\title{Chandra Observations of the Lensing Cluster EMSS 1358+6245: Implications
for Self-Interacting Dark Matter}
\author{J.S.\ Arabadjis\altaffilmark{1}, M.W.\ Bautz\altaffilmark{1}, and
G.P.\ Garmire\altaffilmark{2}}
\altaffiltext{1}{Center for Space Research, Massachusetts Institute of
Technology, Cambridge, MA 02139; {\tt jsa@space.mit.edu},
{\tt mwb@space.mit.edu}}
\altaffiltext{2}{Department of Astronomy \& Astrophysics, 525 Davey Laboratory,
The Pennsylvania State University, University Park, PA 16802}

\begin{abstract} 

We present Chandra observations of EMSS 1358+6245, a relaxed cooling flow
cluster of galaxies at $z=0.328$.  We employ a new
deprojection technique to construct temperature, gas, and dark matter profiles.
We confirm the presence of cool gas in the cluster core, and our
deprojected temperature profile for the hot component is isothermal
over 30 kpc $< r <$ 0.8 Mpc.  Fitting the mass profile to an
NFW model yields $r_s = 153^{+161}_{-83}$ kpc and $c = 8.4^{+3.4}_{-2.3}$.
We find good agreement between our dark matter profile and weak gravitational
lensing measurements.  We place an upper limit of 42 kpc (90\% confidence
limit) on the size of any constant density core.  We compare this result to
recent simulations and place a conservative upper limit on the dark matter
particle scattering cross section of 0.1 cm$^2$ g$^{-1}$.  This limit implies
that the cross-section must be velocity dependent if the relatively shallow
core mass profiles of dwarf galaxies are a direct result of dark matter
self-interaction.

\end{abstract}


\section{Introduction} 

Galaxy clusters are powerful laboratories wherein to test the structure
formation theories and simulations of modern cosmology.  For example, while
enjoying many successes in explaining a large number of observational results
\citep{navarro_six,navarro_seven,moore}, the cold dark matter (CDM) paradigm
appears to be inconsistent with details of the structure of cluster dark matter
halos \citep{spergel} and galactic rotation and density profiles
\citep{spergel,moore}.  The question of whether a CDM universe can produce the
observed mass profiles of galaxy clusters without serious modification is
currently unresolved, although the emerging consensus is that CDM alone
probably fails to produce cluster cores like those observed (e.g.\
\citet{spergel,dave,yoshida}; see \citet{taylor}, however).

The launch of the Chandra X-ray Observatory \citep{weisskopf} has made it
possible to study the mass distribution in the central regions of galaxy
clusters where the discrepancies between CDM models and observations are the
most glaring \citep{moore,firmani}.  The $\sim0.$\arcsec5 spatial
resolution of the Chandra/ACIS-S instrument, which corresponds to about 3 kpc
at $z=0.3$, is now comparable to that of ground-based optical measurements,
(although in practice low-N photon statistics limits the actual resolution
achieved).  For nearby clusters, for sufficiently long exposures, one can
measure the mass profile of a cluster in the inner few kpc \citep{david},
scales which are generally inaccessible to gravitational lensing studies.

The advantage possessed by gravitational lensing studies, however, is that
the measurement is independent of the dynamical state of the gravitating
matter.  To derive a total gravitating mass profile from the baryonic X-ray
emission map one assumes hydrostatic equilibrium \citep{sarazin,
allenfabian_ninetyfour}, which in practice means that one assumes that the
cluster is supported via isotropic thermal pressure (i.e.\
$v_{\rm rot}/\sigma \ll 1$ and $|{\bf B}|^2/nkT \ll 1$), and that no recent
merger event has caused a disruption in the pressure, temperature and density
profiles.  One would expect clusters with fairly circular isophotes, and which
contain cooling flows (which are thought to be disrupted in merging events --
see \citet{allenfabian_ninetyseven}, \citet{allen} and \citet{fabian}), to
merit the simplifying hydrostatic treatment.

EMSS 1358+6245 (a.k.a.\ \mbox{CL 1358+6245}, \mbox{ZwCl 6249}) is such a galaxy
cluster.  Discovered optically by Zwicky \citep{zwicky} and later in X-rays
with the Einstein IPC \citep{luppino}, it is a nearly circular, apparently
relaxed cluster, with an Abell richness class 4 \citep{luppino,hoekstrafranx}
and a central bright cooling flow \citep{bautz,allen}.  Because of its moderate
redshift ($z=0.328$), the entire cluster fits on the ACIS S3 chip, leaving
enough blank field to derive a reliable background estimate.  Figure~\ref{f01}
shows a broad band (0.3-7.0 keV) ACIS S3 image of EMSS 1358+6245, adaptively
Gaussian smoothed on scales of 1 to 10 pixels ($\sim$0.5 to 5\arcsec).  The
contour levels are listed from inner- to outermost; contours associated with
point sources in the field have been omitted.  The presence of the cooling flow
and the regularity of the X-ray isophotes suggest that hydrostatic equilibrium
is a reasonable approximation of the dynamical state of the cluster gas.
Figure~\ref{f02} shows a mosaic of archived HST exposures of the central region
of EMMS 1358+6245 with adaptively-smoothed soft-band (0.3-2.0 keV) X-ray
isophotes superimposed.  Note that the peak of the emission (i.e.\ the central
green dot), at a value of 2.95 photons/(0.\arcsec49)$^2$, appears to be
slightly offset from the position of the central dominant galaxy (CDG).  We
shall return to this point in \S\ref{obs}.

The mass of EMSS 1358+6245 has been measured using both weak
\citep{hoekstrafranx} and strong \citep{franx,allen} gravitational lensing, and
estimated from \ASCA\ and \ROSAT\ observations \citep{bautz,allen}.  These
measurements are roughly consistent, with a mass of about $4\tenup{14}$ \Msun\
enclosed within the 1 Mpc.  In this study, we use a Chandra ACIS-S imaging
spectroscopy to derive a mass profile of EMSS 1358+6245 and compare it to
lensing measurements.

We briefly outline the Chandra observation and data analysis procedure in
\S\ref{obs}.  We describe our deprojection and modelling technique in
\S\ref{dep}, and compare our results to previously published measurements in
\S\ref{comp}.  We then use our profile to constrain possible dark matter
matter candidates in \S\ref{dm}.  Finally we summarize our findings in
\S\ref{sum}.  We assume $\Omega_m=1$, $\Omega_\Lambda=0$, and $H_0=50 h_{50}$
km s$^{-1}$ Mpc$^{-1}$ throughout (1\arcsec\ = 5.79 kpc at $z=0.328$ for
$h_{50}=1$).

\section{Chandra Observations, Astrometry, and Data Analysis} \label{obs} 

EMSS 1358+62 was observed with the Chandra X-ray Observatory for approximately
55 ks on 3-4 September 2000 using the S3 chip on the ACIS detector.  The level
2-processed data were aspect-corrected and filtered for periods of high
background using the CIAO software package, according to the standard
procedures described in the \citet{ciaothreads}.

The center of the projected X-ray emissivity (Figure~\ref{f02}) is offset from
the optical position of the CDG by about 2\arcsec\ (about 12 kpc).  The offset,
if real, has important consequences for the dynamical state of the cluster, so
we carefully examined the the astrometry of the Chandra field.  We ran the CIAO
wavelet source detection routine {\it wavdetect} \citep{ciaothreads} on the
ACIS S2/S3 fields (i.e.\ those chips closest to the ACIS-S aimpoint), and
cross-correlated the output with the Tycho-2 \citep{hogA} and USNO-A2.0
\citep{monet} catalogues.  Five USNO sources and zero Tycho sources were within
3\arcsec\ of a wavlet-detected Chandra source.  Figure~\ref{f03} shows the
location of the five USNO sources in relation to the Chandra ACIS S2/S3 field,
the HST mosaic, and the CDG.  Although the Tycho-2 catalogue is nominally more
accurate than the USNO-A2.0 catalogue, with internal astrometric standard errors
in general less than 90 mas \citep{hogB}, above $\delta=-20$ the USNO positions
are usually within 20 mas of the Tycho-2 positions \citep{assafin}, making
USNO-A2.0 as effective an astrometric standard.  Of these five sources, two
appear somewhat extended, either because they are galaxies or because they
are sufficiently far from the ACIS aim point.  The remaining three point
sources were used as astrometric anchors to calculate the difference between
the Chandra and USNO coordinates.  These offsets are shown as filled circles in
Figure~\ref{f04}.  The mean offset is represented by the small open circle, and
its error by the ellipse.  About half of the CDG/X-ray peak offset is due to
the Chandra field astrometry.  Figure~\ref{f05} shows the center of the cluster
including this astrometric correction.  The top panel shows broad-band (0.3-7.0
keV) X-ray contours superimposed on the 0-10 pixel adaptively smoothed,
$2\times2$-binned broad-band Chandra image.  The same contours are also shown
on the unsmoothed $2\times2$-binned Chandra image (middle) and on the central
field of the HST mosaic (bottom).  The offset between the CDG and the X-ray
peak that remains after the astrometric correction is 0.\arcsec99, while the
error in the correction is 0.\arcsec97.  Therefore, the CDG and the X-ray
emission peak are cospatial at about the 1-$\sigma$ level.

After locating the center of the projected emissivity we divided the cluster
into 10 concentric annuli such that each contained a minimum of 2000 counts
(1600-2000 after background subtraction), the center annulus being a disk.
Owing to decreased signal-to-noise in the outer bins, annuli 9 and 10 were
constructed to contain roughly 4000 (2700) and 10,000 (3300) photons,
respectively.  We found that the extra counts in the last two annuli were
required for numerical stability during the iterative fitting procedure, and
they also helped to reduce the ``noise'' in the derived temperature profile.
The annuli ranged (outer radius) from 11.5 to 285.8 pixels, which translates to
5.66 to 140.56 arcsec.  At $z=0.328$, this corresponds to $\sim30$ to 800 kpc,
for $H_0=50$ km s$^{-1}$ Mpc$^{-1}$ and $q_0=0.5$ (see Table~\ref{t01}).

All obvious point sources were removed, and a spectrum was extracted from each
annulus.  RMF and ARF response matrices were constructed according to the
\citet{ciaothreads}, as was a background spectrum from an annulus exterior to
the outermost bin.  The spectra were recorded in PI format, and grouped such
that there were a minimum of 20 counts per channel.

\section{Cluster Deprojection and Mass Determination} \label{dep} 

We assume that the cluster is a spherical, self-gravitating pressure-supported
plasma whose X-ray emission is optically thin.  We assume hydrostatic
equilibrium throughout (except for small blobs of cold gas in the innermost
regions which, although not hydrostatic, are assumed to be in a steady state).
The hydrostatic equation can be written \citep{sarazin}
\begin{equation}
M(r) = -\frac{kT}{G\mu m_p/r}
\left( \frac{d\log{T}}{d\log{r}} +\frac{d\log{\rho}}{d\log{r}} \right) \, ,
\label{eq01}
\end{equation}

\noindent where $T$ and $\rho$ are the local (baryonic) gas temperature and
density, $r$ is the spherical radius, $M(r)$ is the total mass enclosed
within $r$ (i.e.\ baryons plus dark matter), and $m_p$ and $\mu$ are the
proton mass and mean particle weight, respectively.

To check our assumption of spherical symmetry we examined the surface
brightness map for deviations from circular symmetry.  Specifically, we
performed a simple analysis to estimate the ellipticity $\epsilon$ (and
position angle $\theta$) of the cluster elongation.  We calculate the
luminosity of a pair of wedges A, centered on the X-ray peak of the cluster,
and compare it to a second pair B perpendicular to the first (see
Figure~\ref{f06}).  The wedges have semiwidth $\delta$, radius $R_A$
(or $R_B$), and phase $\theta$ (or $\theta+\pi/2$).  The luminosity of the two
wedge pairs is then

\begin{equation}
L_A(R_A,\theta) = \int_{0}^{R_A} r dr \left[ \,
\int_{\theta-\delta}^{\theta+\delta} \Sigma(\theta,r) \, d\theta +
\int_{\theta+\pi-\delta}^{\theta+\pi+\delta} \Sigma(\theta,r) \, d\theta \,
\right]
\label{eq02}
\end{equation}

\noindent and

\begin{equation}
L_B(R_B,\theta) = \int_{0}^{R_B} r dr \left[ \,
\int_{\theta+\pi/2-\delta}^{\theta+\pi/2+\delta} \Sigma(\theta,r) \, d\theta +
\int_{\theta+3\pi/2-\delta}^{\theta+3\pi/2+\delta} \Sigma(\theta,r) \, d\theta
\, \right]
\label{eq03}
\end{equation}

\noindent Maximizing $L_A(\theta)$ for $R_A/R_B=1$ yields the position angle,
and then scaling $R_B/R_A$ until $L_A/L_B=1$ gives an estimate of the axial
ratio, which is simply $1+\epsilon$.  The noise in the surface brightness map
that tends to inflate $L_A/L_B$ for some $\theta$ is ameliorated somewhat by
the smoothing introduced by the finite window size $\delta$.  Figure~\ref{f07} 
shows the axial ratio and position angle as a function of $R_A$.  The
flattening is consistent with an ellipticity of 0.34 over an order of magnitude
in $R_A$, for $\delta=\pi/36$.  Because much of this flattening is due to
substructure in the soft emission due to the cool component (which shows an
axial ratio of 2.5 for $R_A=200$), we feel confident that the assumption of
spherical symmetry, especially in the hot, pressure-dominating component, is
not unreasonable.  We shall, however, return to the implications of
$\epsilon=0.3$ in \S\ref{dm}.

Se also fit a 2-D $\beta$ model to the cluster emission map using the Sherpa
data analysis package \citep{ciaothreads}.  The profile is described by

\begin{equation}
f(x,y) =
\frac{f_0}{ \left( \, 1 + \frac{r^2}{r_0^2} \, \right)^{3\beta - \frac{1}{2}} }
\label{eq04}
\end{equation}

\noindent where

\begin{equation}
r = \sqrt{ u^2 + v^2/(1-\epsilon)^2 } \, \, ,
\label{eq05}
\end{equation}

\noindent $u$ and $v$ are the rotated coordinates:

\begin{equation}
\left(
\begin{array}{c}
u\\v
\end{array}
\right) =
\left(
\begin{array}{ccc}
\phantom{-}\cos{\theta} & + & \sin{\theta} \\
-\sin{\theta} & + & \cos{\theta}
\end{array}
\right)
\left(
\begin{array}{c}
x\\y
\end{array}
\right) \, \, ,
\label{eq06}
\end{equation}

\noindent and $\epsilon$ and $\theta$ are the isophotal ellipticity and
position angle, respectively.  (We have assigned $(x,y)=0$ to the X-ray
emission peak.)  Here $r_0$ is the core radius and $f_0$ is the central surface
brightness.  We fit for the core radius, the amplitude, and the isophotal
ellipticity and orientation.  We use the maximum likelihood method of
\citet{cash}, as implemented in Sherpa \citep{ciaothreads}, rather than a
$\chi^2$ minimization, because the vast majority of the image pixels contain
fewer than $\sim20$ photons. 

We find an ellipticity of $0.174\pm0.008$ for the cluster (the position angle
has an oblate/prolate degeneracy), although the uncertainty is underestimated
due to parameter correlations.  This indicates again that spherical symmetry is
a reasonable approximation.  The best fit parameter values are $r_0=130\pm30$
kpc and $\beta=0.69\pm0.02$.

In order to derive {\it spherical} radial profiles we construct a model
consisting of $N$ concentric spherical shells whose inner and outer radii
correspond to the inner and outer cylindrical radii of the projected annuli in
the data set.  (Hereafter annuli/shells will be labeled 1 through N, in order
of increasing radius.)  Quantities can be mapped between the two geometries
through the upper diagonal matrix ${\sf V}$, whose elements $V_{ij}$ contain
the volume of spherical shell $j$ intersected by a cylindrical shell formed by
the projection of annulus $i$.  If the volume emissivity of the gas in
spherical shell $j$ is $u_j$, then the luminosity of annulus $i$, $H_i$, is
simply

\begin{equation}
{\bf H} = {\sf V} \cdot {\bf u} \, \, ,
\label{eq07}
\end{equation}

\noindent where we have written the relation for all annuli simultaneously.
Since $H$ is related to the surface brightness $I$ through
$H_j = 2\pi (r_j^2 - r_{j-1}^2) I$, the deprojection of the surface brightness
is accomplished by inverting ${\sf V}$, which is completely specified by the
binning geometry.  Because ${\sf V}$ is upper diagonal, the solution set of
volume emissivities is easily obtained working from the outer- to the innermost
annulus.  The emissivity of the outermost shell is simply $u_N = H_N/V_{NN}$,
while the rest of the shells are calculated moving inward using

\begin{equation}
u_j = \frac{2\pi (r_j^2 - r_{j-1}^2) I_j - {\displaystyle \sum^{N}_{i=j+1}} \,
V_{ji} \cdot u_i}{V_{jj}}
\label{eq08}
\end{equation}

The volume emissivity of each spherical shell is characterized by a temperature
and a normalization.  The X-ray emission from each shell is modelled
spectroscopically using MEKAL \citep{meweg,mewel,kaastra,liedahl} model in the
XSPEC software package \citep{arnaud}, which describes the emission from an
optically thin thermal plasma.

Since a strong cooling flow signature was seen in both ASCA and ROSAT
observations \citep{bautz,allen}, we first sought to verify its existence in
the Chandra data.  We modelled the inner two annuli separately using the
cooling flow model CFLOW of \citet{mushotzky} in XSPEC, together with a MEKAL
component and an intervening absorber.  The cooling flow model is described by
a mass deposition rate $\dot{M}$, a hot gas temperature and a low-temperature
cut-off.  We assume that the flow cools from the ambient cluster gas, so we
pinned the hot gas temperature to the MEKAL temperature during the fitting
procedure.  We also assume that the emissivity is proportional to the inverse
cooling time at the local temperature.  An acceptable fit
($\chi^2$/d.o.f.\ = 62.37/69) was obtained for a mass deposition rate of
$40.1^{+6.9}_{-18.9}$ \Msun\ y$^{-1}$ (90\% confidence interval) and a low-$T$
cut-off $\sim1$ keV.  This result differs significantly from that of
\citet{allen}, who finds $\dot{M} = 690^{+350}_{-290}$ \Msun/y.  The
discrepancy is due to the extra (internal) absorption component which that
study assigns to the cooling flow emission.  While we find no particularly
compelling reason to incorporate an additional absorber in this type of model
\citep{arabadjis}, we do not argue this point in the present study, since we
merely wish to establish the presence of a cool emission component in the
Chandra data.

The bulk of the emission in the flow originates from the coolest gas, and so
in the interest of simplicity we modelled the cooling flow component in the
complete cluster analysis as a second MEKAL component, rather than with the
CFLOW model.  To determine the appropriate spatial extent of the cool gas
we tried adding a second emission component (at lower temperature) to several
of the inner spherical shells.  (We shall refer to these two temperatures as
$T_h$ and $T_c$, for the hot cluster gas and the cooling flow gas,
respectively.)  We ran models with two components in (A) shell 1, (B) shells 1
and 2, (C) shells 1 through 3, and (D) 1 through 4.  Model B had a lower
reduced chi-squared value than did model A, and the temperature and
normalizations of cool components in models C and D were essentially
unconstrained.  We therefore adopted B as our working model, and note that this
suggests that the cool gas extends out to about 70 kpc.  It should also be
noted that, unlike other deprojection methods, this method makes no assumptions
about the shape of the temperature and density profiles other than spherical
symmetry and the finite width of radial bins.

The number of concentric shells that can used to construct this model is
limited by two quantities, the memory allocation of the XSPEC program and the
number of photons above the background in the spectrum.  An active XSPEC model
is limited to 1000 parameters, with only 100 allowed to vary at any time.  Each
MEKAL component contains 6 parameters, but we allow only $kT$ and the
normalization to vary.  In addition, we include an intervening column of
Galactic material which absorbs all emission components equally.  For $N$
concentric annuli in the data, there are $N+N_2$ emission components, where
$N_2$ is the number of two-component shells.  Including the absorbing column
yields a model with $6(N+N_2)+1$ parameters.  This model is applied to $N$ data
sets, with the normalizations across them scaled by the appropriate geometric
factors (i.e.\ ratios of ${\sf V}$ terms).  Thus the complete XSPEC model
contains $N[6(N+N_2)+1]$ components, with $2(N+N_2)+1$ of them variable.  In
our case we had $N_2=2$ two-component shells, thus limiting $N$ to 11.

In practice, however, we found that we had to limit $N$ to 10 because the
finite number of photons in our 55 ks spectrum caused instability during the
numerical iteration if the temperature of any given shell was sufficiently
unconstrained.  It turns out that $\gtrsim 1800$ source photons per annulus are
required to bring the ``noise'' in the fitted temperature profile to a level
where numerical stability is achieved throughout the iterative process, which
in our case resulted in 10 annuli (using $N_2=2$).  Thus our model contains a
total of 730 parameters, with 25 of these floating, well under the XSPEC limit.

The best-fit model has a chi-squared of 942.4 for 891 degrees of freedom (there
are 916 spectral bins in the data).  The temperature and density profiles are
shown in Figure~\ref{f08}.   The model shown at left allows the temperature of
each shell to vary.  Note that the temperature profile is consistent with
isothermality, with an error-weighted average temperature of  $7.16\pm0.10$
keV, while the right side pins all the shells at 7 keV.  In each case the
second emission component is shown as the pair of points near $R =$ 0.1 and
0.2 arcsec.  The inferred density profiles are nearly indistinguishable.

The temperature and density profiles can be used to derive an entropy profile
of the cluster.  The discovery of an ``entropy floor'' in galaxy clusters
\citep{ponman} and its subsequent measurement in many clusters
\citep{lloyddavies} have led groups to attempt to explain it in terms of an
additional source of heat (SNe and/or AGN activity) which acts either to
preheat the gas that collapses to form the cluster, or which heats the cluster
throughout its formation \citep{balogh,loewenstein,wu}.  The specific entropy
of the cluster baryons can be calculated using

\begin{equation}
S = T n_e^{-2/3}
\label{eq09}
\end{equation}

\noindent where $n_e$ is the electron number density of the gas.  In
Figure~\ref{f09} we plot the entropy profile of the cluster.  The cold gas
entropy is shown for the inner two shells as well.  The entropy of the hot
cluster gas toward the center is consistent with the entropy floor of 70-140
keV cm$^2$ reported by \citet{lloyddavies} in their sample of 20 clusters.

The temperature and density profiles can also be used according to
equation~\ref{eq01} to measure the total gravitating mass in the cluster.
Figure~\ref{f10} shows the mass profiles corresponding to the
temperature/density profiles of Figure~\ref{f08}, with the logarithmic
derivatives calculated as simple differences.  The data points in the left
panel represent $M(r)$, the mass contained within spherical radius $r$,
calculated using the density and unrestricted temperature profile shown in the
left panels of Figure~\ref{f08}.  The solid blue curve represents the best-fit
``universal density profile'' of \citet{navarro_six,navarro_seven}, hereafter
the NFW profile, a two-parameter family of models whose functional form is

\begin{equation}
\frac{\rho(r)}{\rho_0} = \left( \frac{r}{r_s}\right)^{-1}
                                       \left( 1+\frac{r}{r_s}\right)^{-2}
\label{eq10}
\end{equation}

\noindent where $r_s$ is a scale length,
$\rho_0 =  \delta_c \, \rho_{crit}(z)$, and $\delta_c$ is a characteristic
(dimensionless) density dependent upon the formation epoch of the dark matter
halo.  $\rho_{crit}(z)$ is the critical density at the observed redshift;
for a matter-dominated, $\Omega_0=1$, $\Lambda=0$ universe this is

\begin{equation}
\rho_{crit} = \frac{3H_0^2}{8\pi G} \, (1+z)^3
\label{eq11}
\end{equation}

\noindent
The total mass scales as $h^{-1}$, so our baryon fraction is weakly dependent
upon $h$.  Of course, radial distances such as $r_s$ scale as $h^{-1}$.

We report our best-fit values as $r_s$ and $c$, the concentration parameter,
defined by

\begin{equation}
\delta_c = \frac{\rho_0}{\rho_{crit}} = \frac{200}{3} \,
\frac{c^3}{\ln{(1+c)}-c/(1+c)}
\label{eq12}
\end{equation}

In practice, because the form of our mass profile is $M(r)$ and not $\rho(r)$,
we use the integrated mass of the NFW profile as our fitting template, rather
than the density:

\begin{equation}
M(r) = M_0 \, [ \, \ln(1+r/r_s) + (1+r/r_s)^{-1} -1 \, ]
\label{eq13}
\end{equation}

\noindent where $M_0 = 4\pi \rho_0 r_s^3$.  Using $M(r)$ rather than $\rho(r)$
spares us one numerical differentiation, which would augment the profile noise
considerably.  We find $r_s = 153^{+161}_{-83}$ kpc and $c = 8.4^{+3.4}_{-2.3}$.

\citet{kelson} have measured the central velocity dispersion of the galaxy
located at the cluster center using the Keck I 10m telescope.  Using the
low-resolution imaging spectrograph with a square 1.05\arcsec aperture, they
infer a central velocity dispersion of 311$\pm$11 km s$^{-1}$.  From HST
photometry they obtain an effective radius of $r=3.9$\arcsec\ (23 kpc).
Asssuming a singular isothermal density profile with an istropic velocity
dispersion, the enclosed mass is given by:

\begin{equation}
M_{iso}(r) = \frac{2\sigma^2 r}{G}
\label{eq14}
\end{equation}

\noindent The dashed black line at left (Figure~\ref{f10}) represents an
isothermal sphere at $\sigma=311$ km s$^{-1}$, with the \citet{kelson}
measurement shown as a filled circle plotted at the effective radius.  Although
the point barely falls within the 1-$\sigma$ envelope, it should be remembered
that this point is a model extrapolation from the aperture radius.  If one
uses the aperture radius rather that the effective radius, the point lies
very near the NFW best-fit line.  We shall return to this point in
\S\ref{dm}.

The solid blue curve at the right half of Figure~\ref{f10} shows the best-fit
($\chi^2/d.o.f. = 2.1/6$) spherical NFW profile projected along the line of
sight, i.e.\ $M(r)$ $\rightarrow$ $M(R)$ (in order to compare with the surface
density maps derived from weak lensing measurements; see \S\ref{comp}).
Integrating equation~\ref{eq10} along the line of sight yields
\citep{bartelmann96}

\begin{eqnarray}
M(R) = 4\pi\rho_0 r_s^3 \, \, \,  \cdot & \left\{
\begin{array}{lr}
  \ln(x/2) +
  \frac{1}{\sqrt{1-x^2}} \, {\rm arctanh}\sqrt{1-x^2} \hspace{50pt} & x<1 \\
  \ln(x/2) +
  1 \hspace{168pt} & x=1 \\
  \ln(x/2) +
  \frac{1}{\sqrt{x^2-1}} \, {\rm arctan}\sqrt{x^2-1} \hspace{50pt} & x>1 \\
\end{array} \right.
\label{eq15}
\end{eqnarray}

\noindent where $x=R/r_s$.

Because of the ``noise'' in the solution, one point in the mass profile is
unphysical.  This is caused by a 1-$\sigma$ jog in the temperature profile at
the seventh annulus.  The (positive) logarithmic temperature gradient between
annuli 6 and 7 swamps the density gradient (see equation~\ref{eq01}) causing
$M(r)<0$; this is represented in Figure~\ref{f10} by an open dashed box at an
arbitrary ordinate.

In order to ascertain whether this jog represents an unseen feature in the
data or a fundamental flaw in the method, rather than an ordinary statistical
fluctuation, we varied the initial conditions in the fitting procedure to see
if the iterative process would converge to a solution lacking such a
temperature anomaly.  After many attempts we found that we could only move
the jog to a different annulus.  As a more rigorous check on the method, we
analyzed a cluster simulated with MARX that contained $10^5$ photons and no
background (J.\ Houck \& M.\ Wise, private communication).  We derived a mass
profile for this cluster using 12 annuli, and found two unphysical points due
to jogs in the otherwise smooth temperature profile.  We therefore conclude
that the single unphysical point in our mass profile for EMSS 1358+62 is
{\it not} due to undetected peculiarities in the data, but is more likely a
nonexceptional statistical fluctuation.  We note here that in their
deprojection analysis of Hydra A, \citet{david}, who used a smooth parametric
temperature profile, were still forced to effectively smooth across radial bins
in order to reduce the ``noise'' in the mass profile, even though they had
roughly 5 times the number of photons at their disposal.

Nonetheless, we ran the model again, pinning $T_h$ at 7 keV, to see what effect
this would have on the mass profile.  The resulting temperature and density
profiles are shown on the right side of Figure~\ref{f08}.  The effect on the
mass profile, and its NFW fit, is small; the scale length increases from
$\sim 153$ to 197 kpc, and concentration decreases from 8.4 to 7.0.  The
significance of these values will be discussed in \S\ref{dm} below.

The gas mass and total gravitating mass profiles are shown in Figure~\ref{f11},
for the model wherein $T_h$ is allowed to vary.  The $T_h=7$ keV model is
very similar.  The gas mass fraction rises from $\sim$0.025 at 50 kpc to 0.18
at 0.6 Mpc, scaling as $h_{50}^{-1.5}$ \citep{white}.  This is consistent with
the study of \citet{white}, who find baryon fractions of 0.10--0.22 for a
sample of 13 clusters (90\% confidence).  The mass of an isothermal sphere
based on the \citet{kelson} measurement, and its corresponding profile, is also
shown, as a filled circle and a dashed line, respectively.  Although these data
do not allow us to distinguish between the gas, stars, and dark matter
associated with the CDG, they clearly demonstrate that the total galaxy mass
dominates the gas mass within the cluster center.

Our model does not allow us to determine the morphology of the cool gas within
the inner two shells of cluster gas, although from our ellipticity estimation
we suspect that it is fairly aspherical.  However, its mass, density, and
volume are completely specified, since we have assumed pressure equilibrium
between the two temperature components.  For the unconstrained temperature
profile model, roughly 10\% of the core volume is filled with cold gas,
dropping to a tenth this value at the adjacent shell.  Table~\ref{t02} lists
the relevant quantities for both gas components in the inner two shells.  We
note that the broadband (0.3-7.0 keV) X-ray luminosity of the cool gas
component in the inner shells is $4.4\tenup{43}$ erg s$^{-1}$, comparable to
the luminosity of the ``X-ray overluminous'' ellipticals studied by
\citet{vikhlinin}.

\section{Comparison with Lensing Studies} \label{comp} 

In order to assess the robustness of our derived mass profiles, we compare them
to the the weak lensing profile of \citet{hoekstrafranx}.  While gravitational
lensing measurements may suffer from biases introduced by distant, uncorrelated
large-scale structure, the effect on the mass determination is expected to be
small for distant, rich clusters; for example, \citet{hoekstra} finds a typical
1$\sigma$ uncertainty of about 6\% for measurements out to 1.5 $h_{50}^{-1}$
Mpc.  In addition, \citet{allen} has shown that lensing and X-ray mass
determinations are consistent provided one takes into account the presence of
the lower temperature gas in the cooling flow.

Because weak lensing mass profiles measure mass within a {\it projected}
radius, we integrate the best-fit NFW profile (equation~\ref{eq10}) along the
line of sight (equation~\ref{eq15}).  For both our treatments of the
temperature profile of the baryons (i.e.\ variable vs.\ frozen at 7 keV), the
resulting mass profile agrees well with the weak lensing profile.
The strong lensing measurement \citep{franx,allen} at 120 kpc exceeds the X-ray
and weak lensing measurements by a factor of $\sim 1.6$.  We note that although
the lensing results we quote and our X-ray measurement assume spherical
symmetry, \citet{hoekstrafranx} find evidence for an elliptical mass
distribution with an axis ratio of $\sim 0.3$.  Our analysis shows that while
the X-ray surface brightness is very nearly circular at large radii,
the axis ratio in the core could be as small as 0.7.  The assumption of
spherical symmetry  leads to an average overstimate of a factor of 1.6 in
strong-lensing analyses of simulated clusters, which are generally not
spherically symmetric \citep{bartelmann95}.

Following \citet{allenettori} we compute an ``effective velocity dispersion''
of our best-fit mass profile.  Using the NFW halo virial mass in the singular
isothermal profile we obtain
$\sigma \equiv \sqrt{50} H(z) \, r_{s} c = 700^{+203}_{-84}$ km s$^{-1}$.
This is substantially lower than the observed line-of-sight velocity dispersion
of \citet{fisher}, $1027 \pm 50$ km s$^{-1}$.  As noted by Fisher et al., the
optical velocity dispersion is probably inflated by substructure along the
line of sight to the cluster. As expected, our effective velocity dispersion
agrees, within measurement errors, with the value of
$\sigma = 780 \pm 50$ km s$^{-1}$ derived by \citet{hoekstrafranx} by fitting
an isothermal sphere to the weak lensing data.

\section{Constraints on Self-interacting Dark-Matter} \label{dm} 

Much has been made of the failure of CDM simulations to reproduce the observed
structure of dark matter halos.  A particular example of the problem is that
CDM halos are steeper than observed dwarf galaxy halos.  Parameterizing the
core profile as $\rho(r) = \rho_0 (r/r_0)^\alpha$ as $r\rightarrow 0$, CDM
simulations suggest either $\alpha=-1$ \citep{navarro_six,navarro_seven} or
$-1.5$ \citep{moore,fukushige}.  H$\alpha$ observations low-surface brightness
galaxies suggest a flatter profile, $\alpha\equiv -0.5$
\citep{swaters,dalcanton}.  Instead of a triaxial, central cusp, as predicted
by CDM simulations, \citet{tyson} find a spherical $r_s = 35 h^{-1}$ kpc core
in the strong gravitational lens \mbox{CL 0024+1654} using a multiply-imaged
background galaxy (see \citet{broadhurst} and \citet{shapiro}, however).  A
weak lensing study by \citet{smail} also indicates soft cores \mbox{CL 1455+22}
and \mbox{CL 0016+16}.  The literature survey of \citet{firmani} suggests that
the central density of clusters is about 0.02 \Msun\ pc$^{-3}$, regardless of
halo mass, whereas CDM predicts values of $\ge 1$ \Msun\ pc$^{-3}$ for dwarf
galaxies, and larger still for more massive halos.  \citet{spergel},
\citet{firmani}, \citet{dave} and others use these discrepancies to argue for
the existence of self-interacting dark matter (SIDM), which will tend to reduce
both the core profile slope and central density of dark matter halos.

X-ray measurements of cluster dark matter profiles, however, fall somewhat more
in line with the CDM simulations.  \citet{tamura} find a central slope of
$\sim -1.5$
for Abell 1060.  \citet{markevitch} find that NFW provides a good description
of the mass profiles they derive for A2199 and A496 from \ASCA\ and \ROSAT\
observations, over the entire range for which they are able to derive a
temperature profile.  \citet{allenettori} observed A2390 using the ACIS S3
detector on Chandra and derive a mass profile which is consistent with the
NFW profile, and find a concentration $c=4$, as predicted by CDM.
\citet{schmidt} also find adequate agreement between their mass profile of
A1835, derived from a Chandra ACIS-S3 observation, and the NFW profile, and
consistency with weak lensing measurements.  They calculate a scale radius of
640$^{+210}_{-120}$ kpc and a concentration of 4.0$^{+0.54}_{-0.64}$.  They
find a slightly better fit when they model A1835 as a non-singular isothermal
sphere, deriving a core size for the cluster (distinct from the scale radius of
an NFW fit) of $r_c = 65^{+5}_{-10}$ kpc.  \citet{david} determine a central
slope of $\alpha\sim -1.3$ for Hydra A, and find no evidence for a flat, SIDM
dominated core, down to scales of $\sim 40 \, h^{-1}_{50}$ kpc.  However, their
NFW fit to their mass profile results in a concentration $c=12$, a factor of 3
larger than that predicted by CDM.

In general we find that the NFW model provides an adequate fit to our mass
profile.  Like the \citet{david} study, however, our fit to the mass profile of
EMSS 1358+6245 results in a concentration which is somewhat higher than
that predicted by NFW for rich clusters.  The appendix of \citet{navarro_seven}
provides a prescription for calculating cluster parameters at the collapse
(i.e.\ formation) redshift $z_{coll}$, which is a function of cluster mass in
this model.  Using the Einstein-de Sitter simulation with a CDM spectrum
normalization $\sigma_8=0.63$ and $h=0.5$ (the SCDM model of
\citet{navarro_seven}) we calculate a collapse redshift of $z_{coll}=4.9$; it
is doubtful that a cluster this massive could have formed so early.  NFW
predicts a concentration of about 4 for massive clusters, but the scatter in
their suite of simulations could be as high as 2.

One might ask at what level SIDM can be ruled out by current observations.
Recently, \citet{yoshida} simulated cluster-sized halos and found that
relatively small dark matter cross-sections ($\sigma_{dm}$ = 0.1 cm$^2$
g$^{-1}$) produce relatively large (40 $h^{-1}$ kpc) cluster cores.
\citet{dave} simulated galaxy-sized halos and required interaction
cross-sections as large as 5 cm$^2$ g$^{-1}$ to reproduce the shallow central
slopes of galaxy density profiles.  As noted by \citet{dave} and explored by
\citet{firmaniII}, these findings can be reconciled if the dark matter
interaction cross-section is velocity-dependent.  For example, the two
findings are roughly consistent with

\begin{equation}
\sigma_{dm} = \sigma_0 \left(\frac{V}{V_0}\right)^{-a}
\label{eq16}
\end{equation}

\noindent with $\sigma_0 = 1$ cm$^2$ g$^{-1}$, $V_0 = 100$ km s$^{-1}$, and
$a=-1$.  \citet{hennawi}, however, use supermassive black hole demographics to
rule out SIDM cross sections as large as this.  They present a model with
$a=0$ and $\sigma_0 = 0.02$ cm$^2$ g$^{-1}$ that is consistent with observed
supermassive black hole masses.  While their model does not remedy the
galactic-scale dark matter halo problem, they sketch a scenario wherein the
density cusps in such halos can be softened through black hole mergers.

Our mass profile for EMSS 1358+6245 shows no evidence for a flat core on scales
larger than $40 \, h_{50}^{-1}$ kpc.  In Figure~\ref{f12} we show power law
fits to the mass profile as a function of the outermost point used in the fit.
The profile is consistent with $M_r \propto r^{1.1}$, or $\alpha=-1.9$,
somewhat steeper than CDM predictions.  In fact, the inner profile displays
behavior which is the {\it opposite} of what would be required for nearing a
flat core; $\alpha$ should approach, not diverge from, 0 as $r\rightarrow 0$.
(The restriction is even more stringent since our model uses $H_0=50$
km s$^{-1}$ Mpc$^{-1}$.)  We place an upper limit on the core size by modeling
the profile as a softened isothermal sphere:

\begin{equation}
\rho(r) = \frac{\sigma^2}{2\pi G (r^2+r_c^2)} \, ,
\label{eq17}
\end{equation}

\noindent We find a core size $r_c = 12^{+30}_{-12} \, h_{50}^{-1}$ kpc
(90\% confidence; $\chi^2/d.o.f. = 4.2/6$).  The contribution of the CDG to the
mass profile for $r>30$ kpc is sufficiently small (Figure~\ref{f10}) that we
can directly compare our core size limit, $r<42 \, h_{50}^{-1}$ kpc, with
\citet{yoshida}.  The most conservative limit we can place must include the
flattening that we determined in \S\ref{dep}.  Although the cluster as a whole
exhibits an ellipticity of 0.34, it could be higher toward the core (see
Figure~\ref{f07}.  The worst-case scenario would be that our innermost annulus
just grazes the minor axis of a core; in that case our limit should be scaled
upward by a factor of $1+\epsilon$.  Furthermore, in keeping with a
conservative posture, if we assume that the axial ratio in the core (near
$r=35$ kpc; see Figure~\ref{f07}) is as high as 7/4, then our core limit is
74 kpc, allowing us to conservatively rule out SIDM with $\sigma_{dm}\geq 0.1
$ cm$^2$ g$^{-1}$.

In order to strengthen the case for the core size upper limit, we verified
that our deprojection and modelling technique can indeed detect a core if one
is present.  We created a set of five simulated galaxy cluster observations to
which we applied our deprojection and fitting method.  Each simulated cluster
follows a $\beta$ model surface brightness profile, with $\beta=0.69$.  The
$\beta$ model core radii $r_0$ ranged from 6 to 96 ACIS S-3 pixels (34.7 to
556 kpc), increasing by a factor of 2 for each successive cluster.  Otherwise
the simulated clusters are similar to EMSS 1358+6245 -- $z=0.328$, and the
photons are sampled from a distribution at $T = 7$ keV.  In addition, each of
the 10 annuli contains $\sim 2000$ photons.  Once we have fit for the
temperature and gas density of each shell, the mass profile computed using the
hydrostatic equation, which is then fit using a non-singular isothermal sphere.
The resulting fits, shown in Figure~\ref{f13}, demonstrate that we can detect
core structures in the mass profile.  (The agreement between $r_0$ and $r_c$
might at first seem too good, particularly when one considers that $r_0$ is a
projected beta model core radius, while $r_c$ is a spherical softened
isothermal sphere core radius.  However, we are pinning down the shape of the
mass profile near the core radius, where the non-singular isothermal sphere is
actually a good approximation to a beta model.  At $r = r_c$, for $\beta=0.69$,
the agreement is better than 5\%.)

In Figure~\ref{f14} we plot the SIDM cross section constraints of several
groups.  The hatch marks delineate the region excluded by each constraint.  The
parameter space external to the triangular region is ruled out by
\citet{hennawi} by requiring that SIDM (1) removes the halo cusp in dwarf
galaxies, (2) does not cause core collapse in 20 km s$^{-1}$ halos, and (3)
does not cause galactic-mass halos to evaporate in a Hubble time.
\citet{hennawi} rule out other areas not shown here by considering the
formation of supermassive black holes (SMBHs), specifically the lack of a
bulge/SMBH in M33, and the presence of a bulge/SMBH in the Milky Way.  They
arrive at SIDM which is unable to remedy the cuspy halo problem of dwarf
galaxies, and and we direct the interested reader to that study.  We have
indicated the results of computer simulations of \citet{yoshida} and
\cite{dave} (see arrows).  Our mass profile of EMSS 1358+6245, in conjunction
with the \citet{yoshida} simulations, provide the constraint
\mbox{$\sigma_{dm}(1000) < 0.1$ cm$^2$ g$^{-1}$}, ruling out the space
indicated in blue.  We observe that the core collapse and halo evaporation
constraints of \citet{hennawi} appear to rule out the model of \citet{dave} in
which SIDM is responsible for dwarf galaxy halo cores.  We note in passing that
our observations are consistent with, though less restrictive than, the
constraints derived by \citet{hennawi} from SMBH demographics.

While our central profile is consistent with NFW, it is intriguing that the
average density within the inner 50 kpc of our mass profile is $\sim 0.025$
\Msun\ pc$^{-3}$, consistent with the \citet{firmani} sample.  If the
\citet{kelson} measurement of the (assumed isotropic) velocity dispersion at
0.55\arcsec\ is extrapolated to $r_e = 3.9$\arcsec\ (22 kpc) as described in
section \S\ref{dep}, the resulting density is, perhaps coincidentally,
$\rho \sim 0.022\pm0.005$ \Msun\ pc$^{-3}$, suggesting that the data allow the
density profile to be flat from 20 to 50 kpc.  It is important to remember,
however, that this extrapolation assumes that it is appropriate to characterize
the CDG density profile as a singular isothermal sphere over the range
$1.05\arcsec \leq R \leq 3.91\arcsec$.  (More conservatively, if the isothermal
model is assumed to hold only at the aperture radius, the profile is consistent
with the absence of a core.)

However uncertain, this result is interesting because it is contrary to what
would be expected from a dark halo associated with the central galaxy, and may
offer a clue to the existence of SIDM which flattens cluster cores on scales of
40 kpc.  The issue can be addressed with a deeper Chandra observation to
enable one to probe the mass profile in X-rays down to the effective radius of
the \citet{kelson} measurement, or through similar observations of other
relaxed clusters.

On the other hand, we note that dark halos of CDGs would steepen the central
density profile and possibly complicate detection of a very small, flat SIDM
core.  Ironically, the relaxed nature of
rich clusters, which allows us to measure their masses by assuming
hydrostatic equilibrium, may turn out to obfuscate the details of the dark
matter distribution.  Even so, Chandra mass profiles appear to justify
high-resolution N-body simulations with very low ($\sigma < 0.1$ cm$^2$ g${-1}$)
dark matter interaction cross sections.

\section{Summary} \label{sum} 

We have used a new spectral deprojection technique to derive a dark matter
profile from a Chandra observation of EMSS 1358+6245.  Our mass profile is
consistent with optical weak lensing measurements \citep{hoekstrafranx}.  It is
nominally consistent with an NFW profile, although our derived concentration
larger than expected from CDM simulations by a factor of a few.  Our best-fit
NFW profile is characterized by $r_s = 153^{+161}_{-83}$ kpc and
$c = 8.4^{+3.4}_{-2.3}$.  We also model the cluster as a non-singular
isothermal sphere and place an upper limit on the core size
$r_c < 42 \, h_{50}^{-1}$ kpc (90\% confidence).  Comparing this value to the
simulations of \citet{yoshida}, we rule out self-interacting dark matter with
cross sections $\sigma_{dm}\geq 0.1$ cm$^2$ g$^{-1}$.

JSA and MWB would like to thank Paul Steinhardt and Marie Machacek for useful
discussions of dark matter candidates and self-interacting dark matter, and
Claude Canizares for proofreading this manuscript.  JSA would like to thank
Joe Hennawi and Jerry Ostriker for a pre-preprint version of their paper.




\clearpage
\plotone{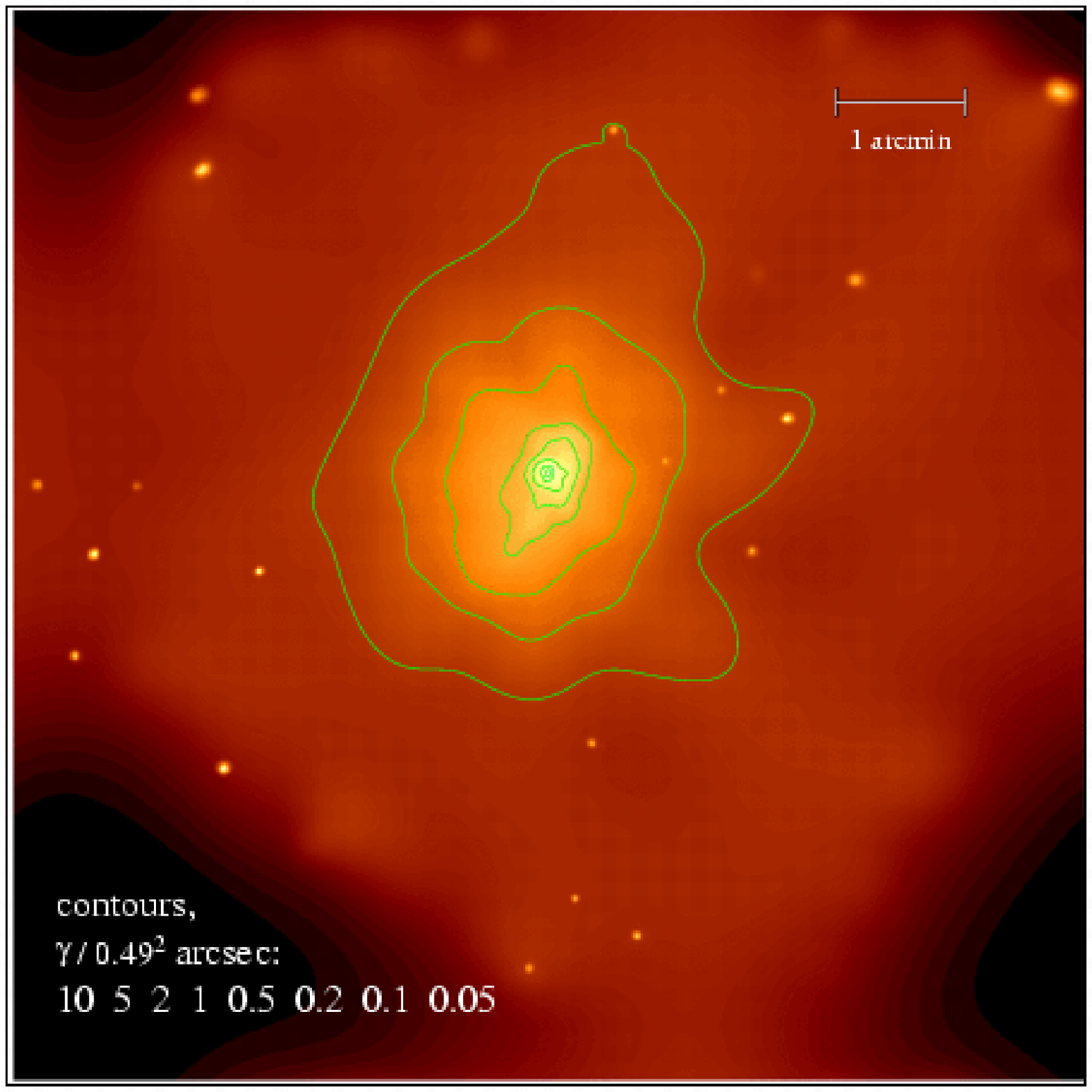}

\figcaption{Broadband ($0.3\leq E/{\rm keV} \leq 7.0$) image of EMSS1358+6245.
\label{f01}}


\clearpage
\plotone{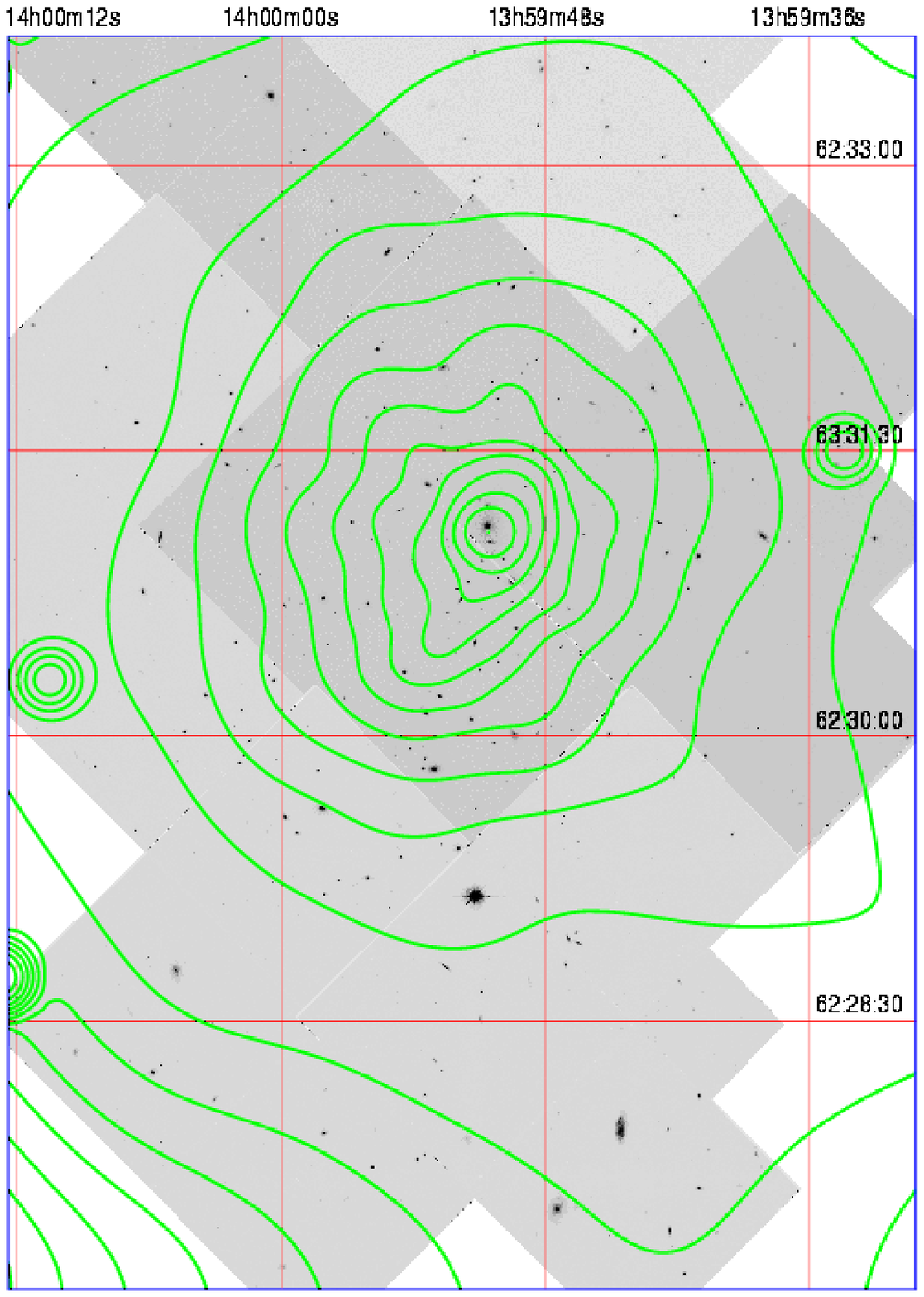}

\figcaption{An HST image of EMSS 1358+6245 with superimposed soft X-ray
($0.3\leq E \leq 2.0$ keV) contours.  Each contour level is 3/5 of the
previous value.
\label{f02}}


\clearpage
\plotone{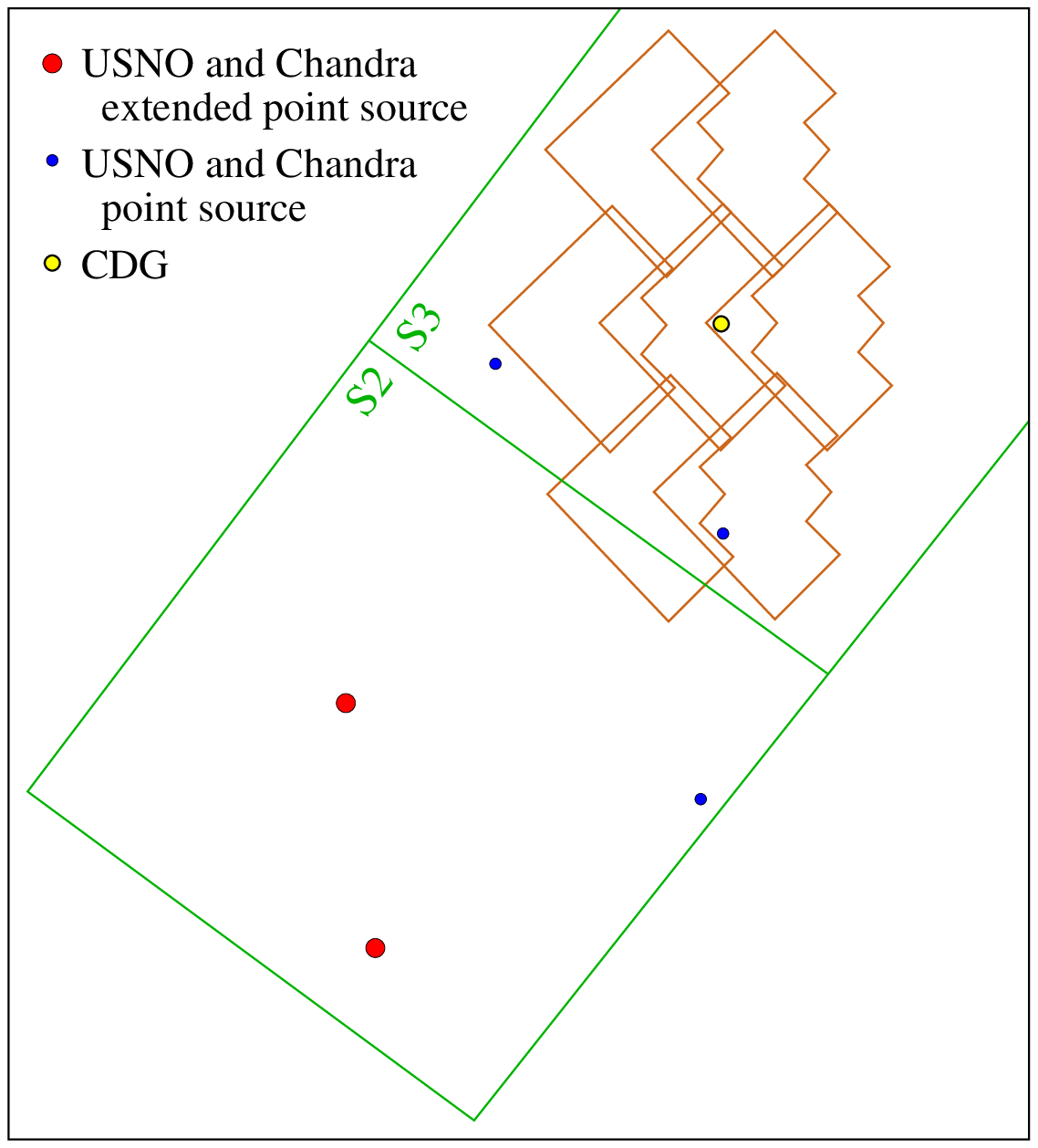}

\figcaption{Five USNO sources in the Chandra ACIS S2/S3 and HST fields.
\label{f03}}


\clearpage
\plotone{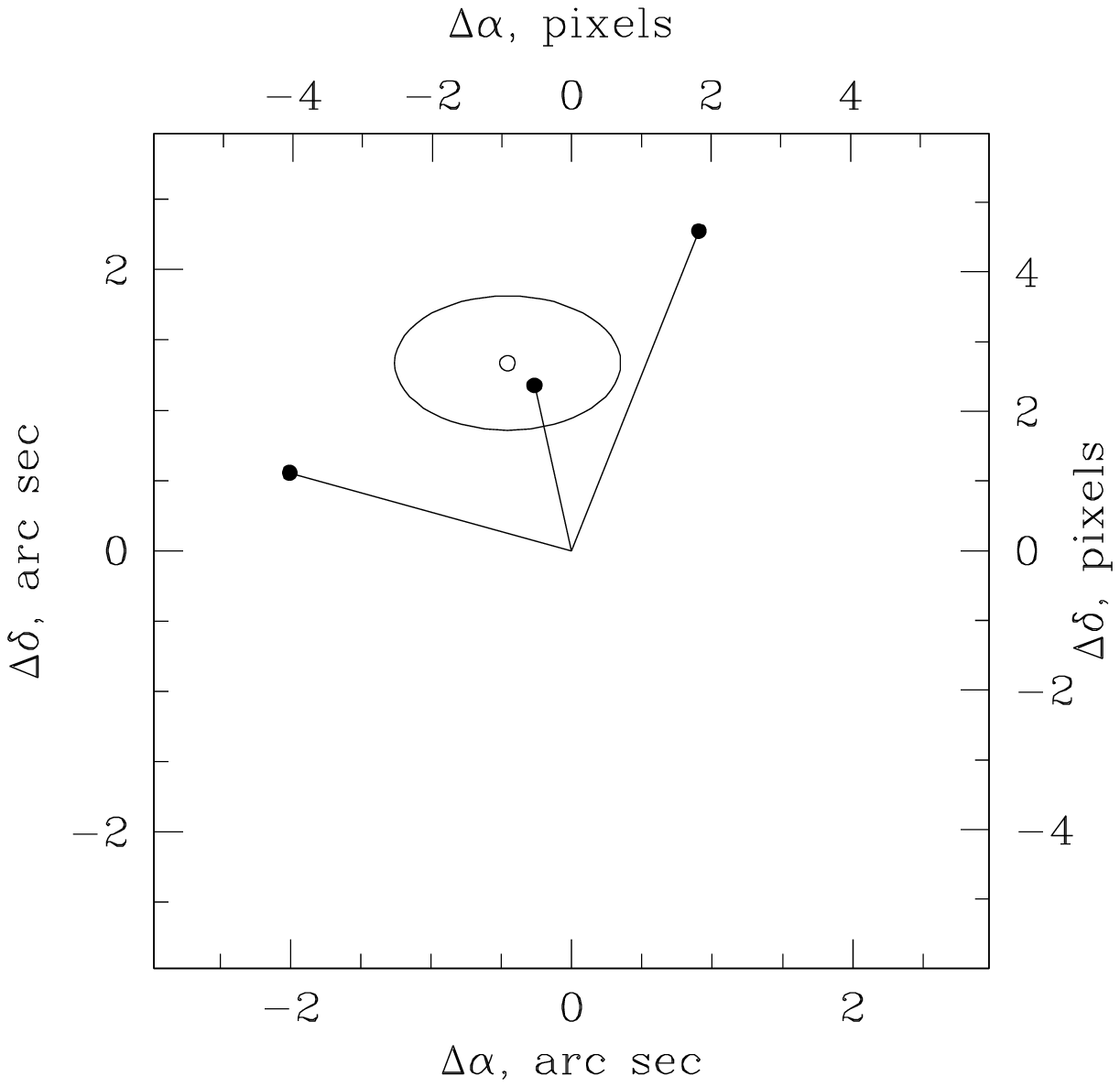}

\figcaption{Positional offsets of the three USNO sources which appear as point
sources on the Chandra fields.
\label{f04}}


\clearpage
\plotone{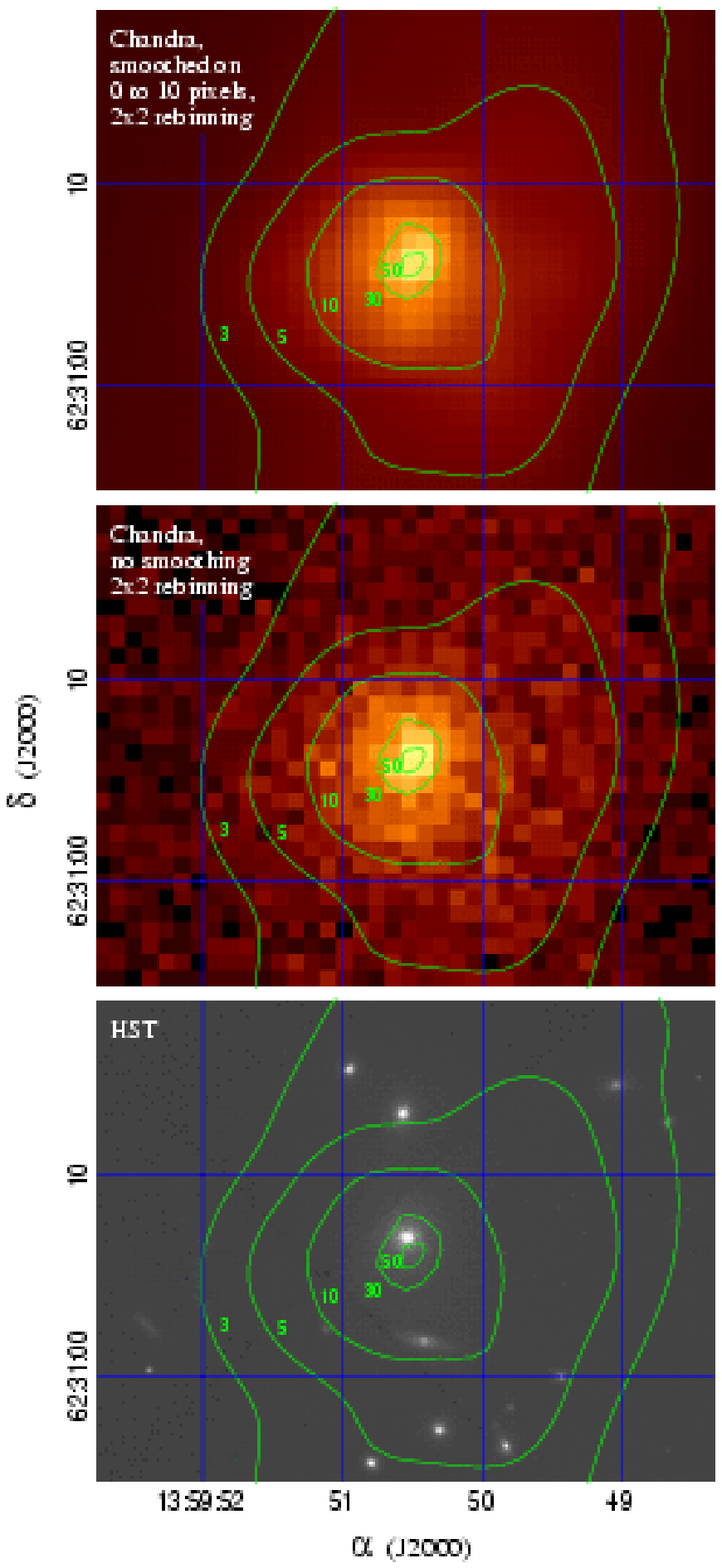}

\figcaption{Broad-band X-ray contours overlaid on the adaptively smoothed
Chandra X-ray image (top), the unsmoothed $2\times2$-binned image, and the HST
field (bottom).
\label{f05}}


\clearpage
\plotone{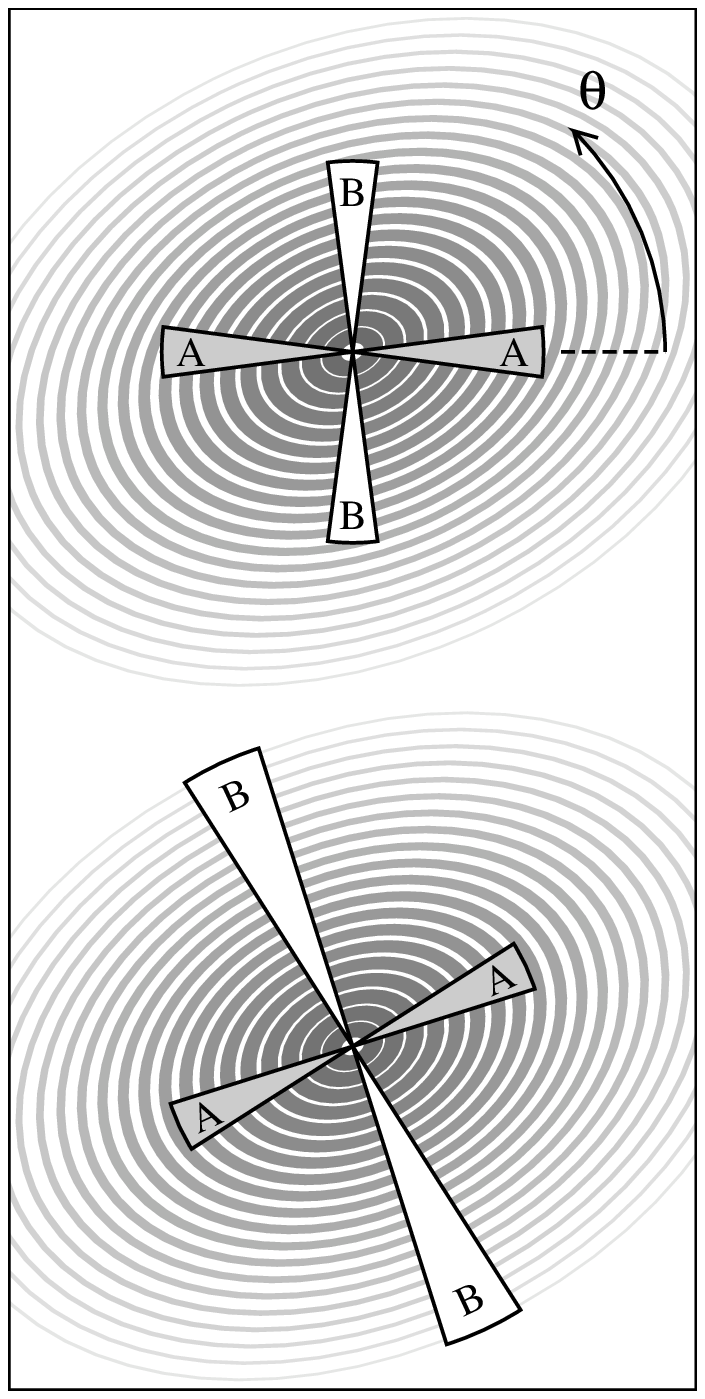}

\figcaption{Wedge pairs used to estimate the cluster ellipticity.  Wedges of
equal radius are used first to determine the position angle, and then wedge
B is scaled so that its luminosity is the same as wedge A.
\label{f06}}


\clearpage
\plotone{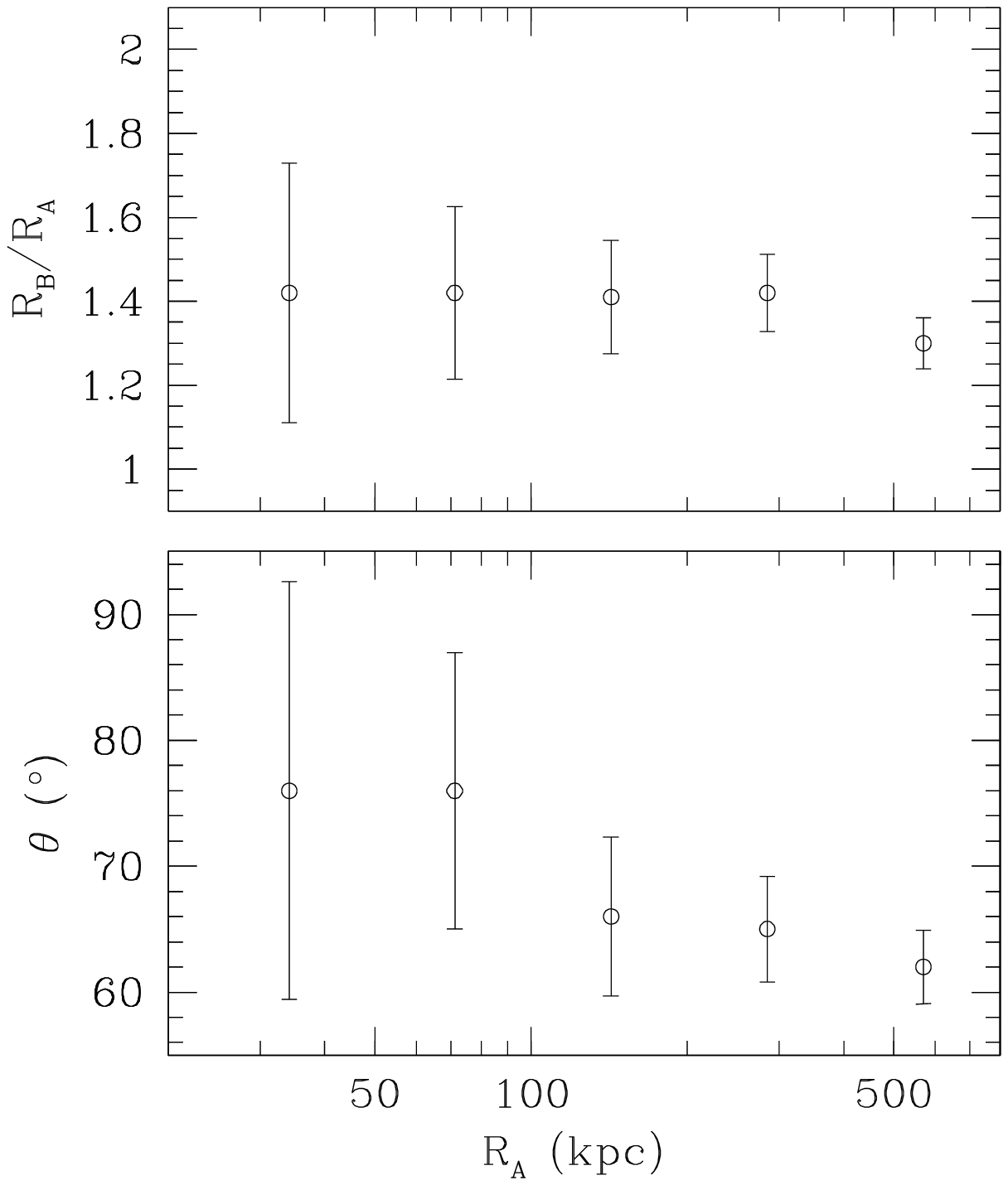}

\figcaption{Radial dependence of the flattening and position angle.
It should be noted that these quantities are calculated by integrating
to $R_A$, and in that sense represent a radially cumulative average.
\label{f07}}


\clearpage
\plotone{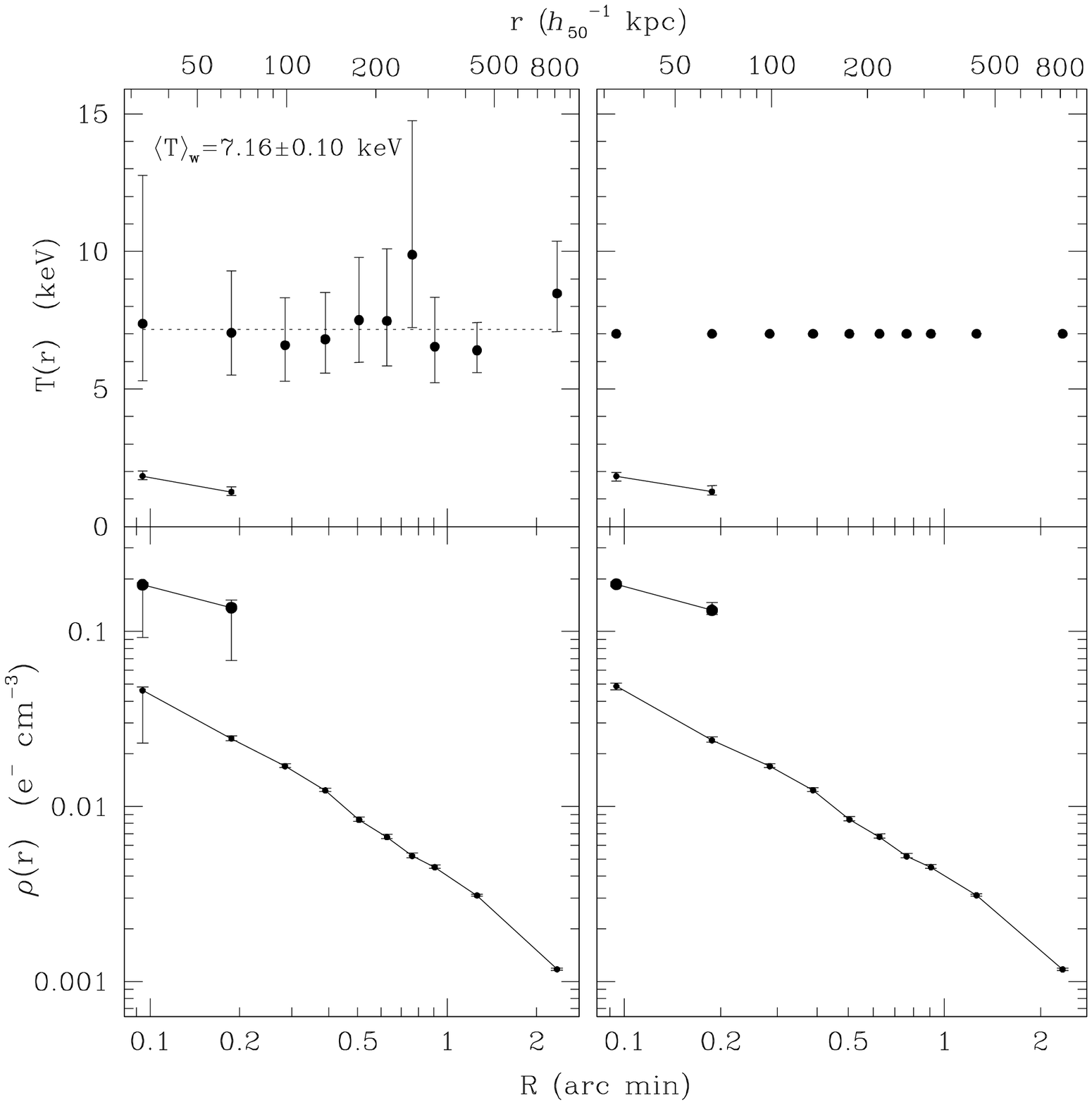}

\figcaption{Deprojected radial temperature and density profiles of EMSS 1358.
Note that the inner two annuli are represented by two emission components at
different temperatures.  The profiles on the left are calculated allowing the
hot gas temperature to vary; those on the right are frozen at 7 keV.
\label{f08}}


\clearpage
\plotone{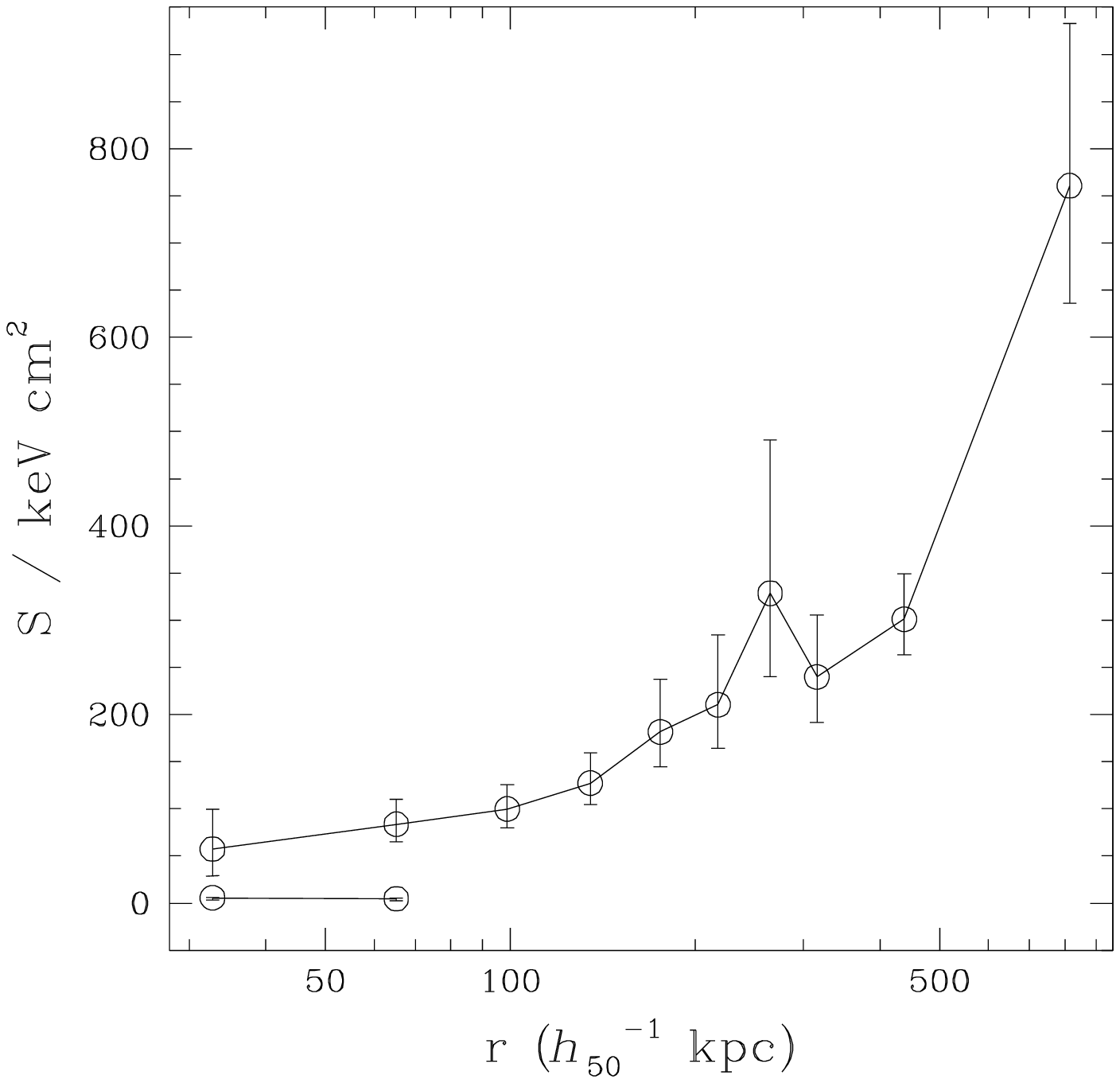}

\figcaption{Entropy profile of EMSS 1358.  The curve that spans the entire
range in $r$ represent the entropy of the gas at temperature $T_h$; the short
segment represents the gas at $T_c$.
\label{f09}}


\clearpage
\plotone{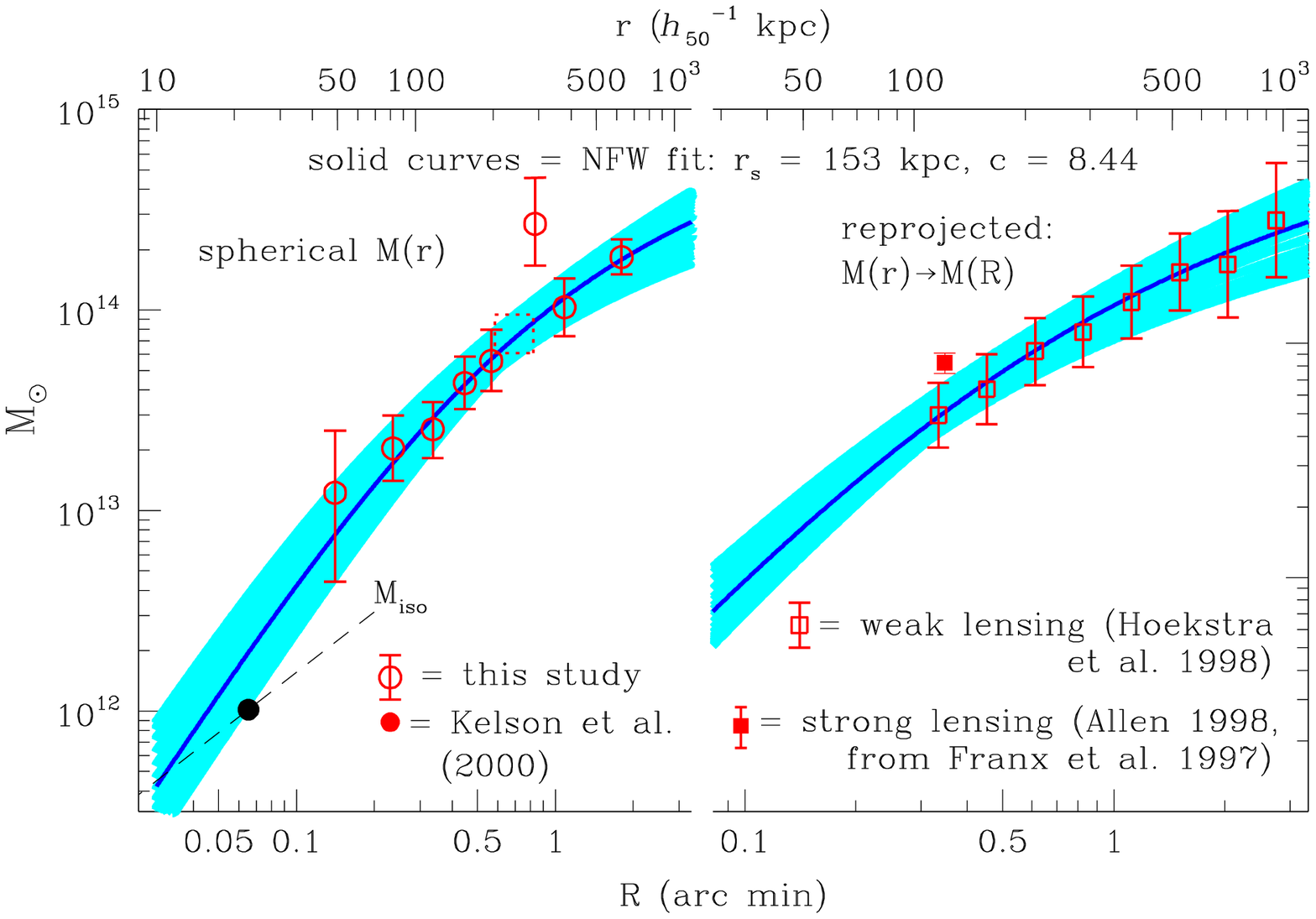}

\figcaption{Deprojected mass profile of EMSS 1358.  The data points in the left
panel represent $M(r)$, the mass contained within spherical radius $r$,
calculated using the density and unrestricted temperature profile shown in the
left panels of Figure~\ref{f08}, with the single unphysical point represented
as a dashed box.  The solid dark blue curve and light blue 1-$\sigma$ envelope
represents the best-fit NFW profile.  On the right the spherical NFW profile
has been projected along the line of sight according to equation~\ref{eq15}, in
order to compare it to the weak lensing measurements of \citet{hoekstrafranx}.
\label{f10}}

\clearpage
\plotone{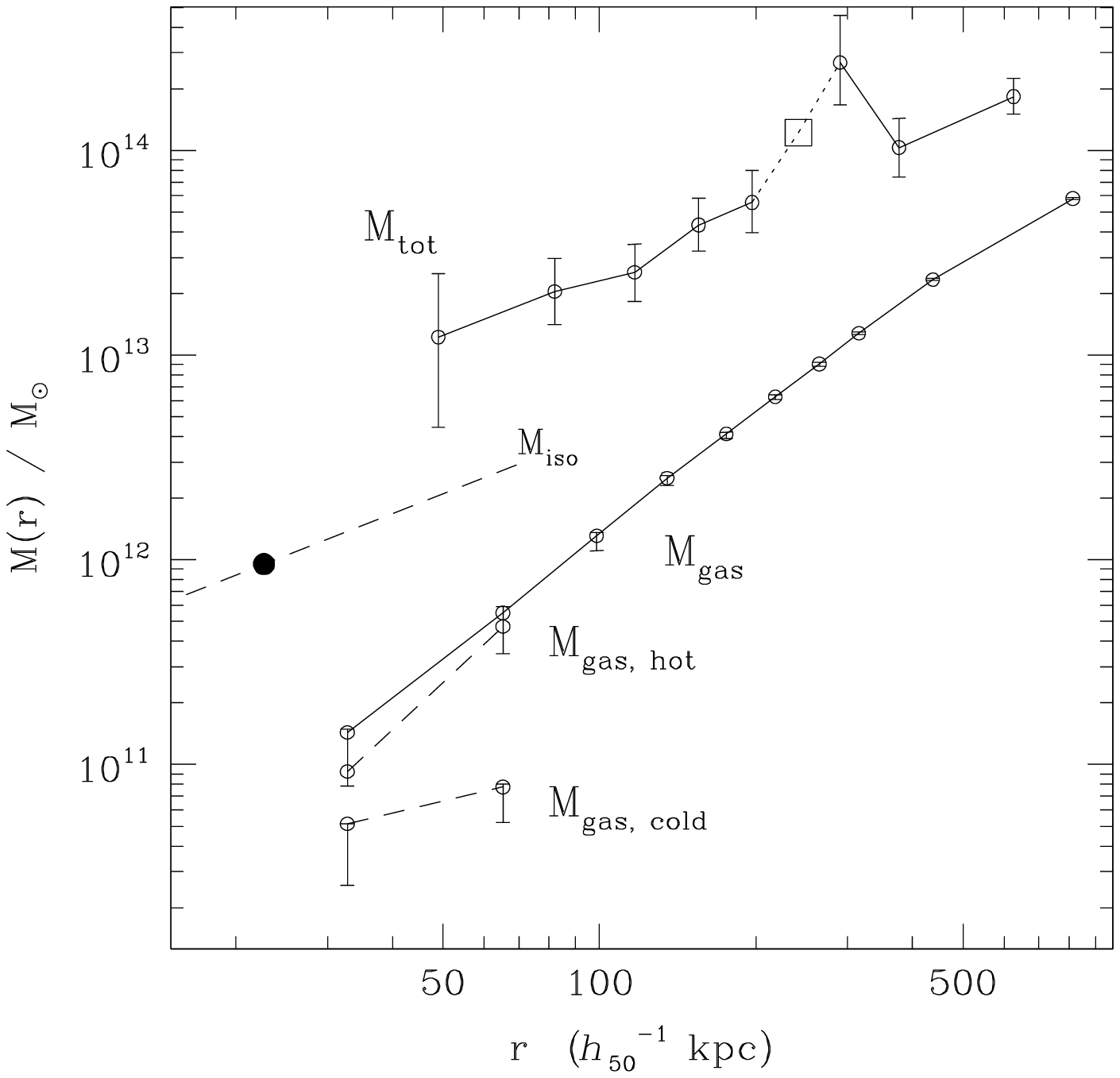}

\figcaption{The total integrated gravitating mass and gas mass of EMSS 1358.
The non-physical point ($M(r) < 0$) is represented by an open box in the total
mass profile.
\label{f11}}

\clearpage
\plotone{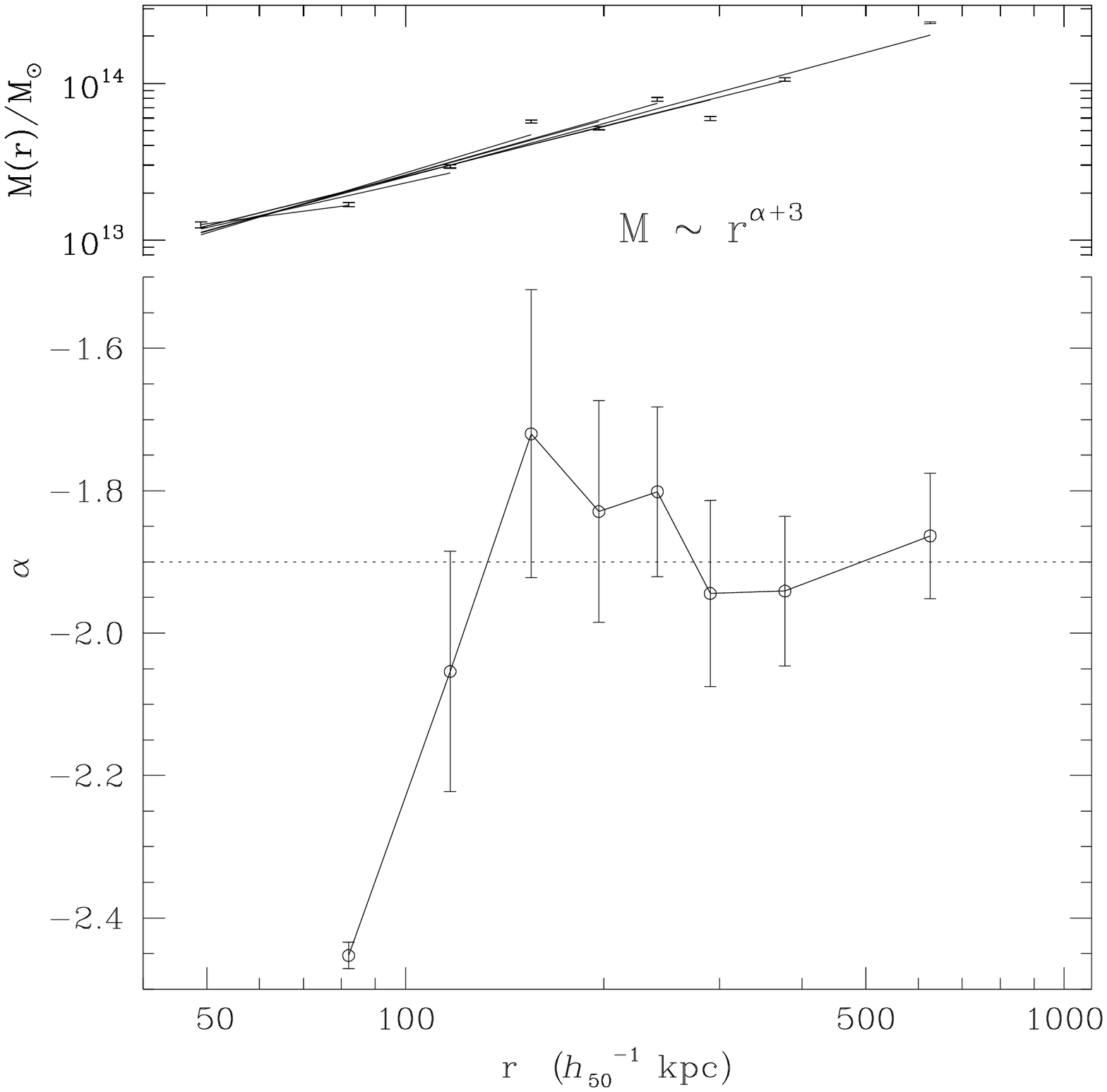}

\figcaption{Power law fits to the mass profile of EMSS 1358.  The top panel
shows the mass profile for the $T_h=7$ keV model along with power law fits
using points 1-$k$, $k\in[2,9]$.  The bottom panel shows the power law
index of the fit, as a function of the radius of the outermost point used.
Note that $\alpha$ is the power law slope on $\rho(r)$, {\it not} $M(r)$.
\label{f12}}

\clearpage
\plotone{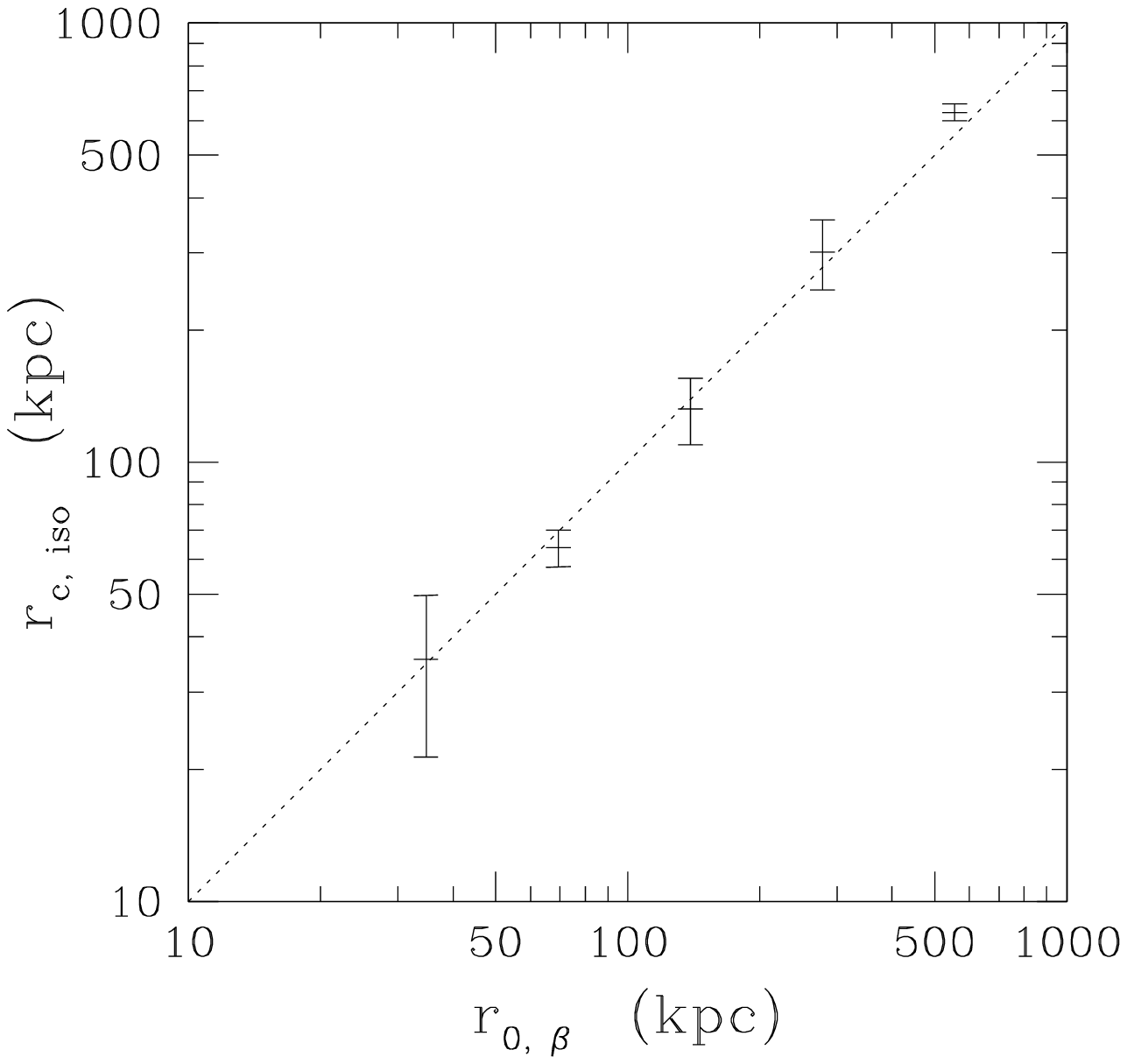}

\figcaption{Softened isothermal sphere fits to a set of 5 simulated galaxy
cluster observations which follow a $\beta$ model profile.  The fitted
isothermal sphere core radius is plotted against the $\beta$ model core
radius.
\label{f13}}

\clearpage
\plotone{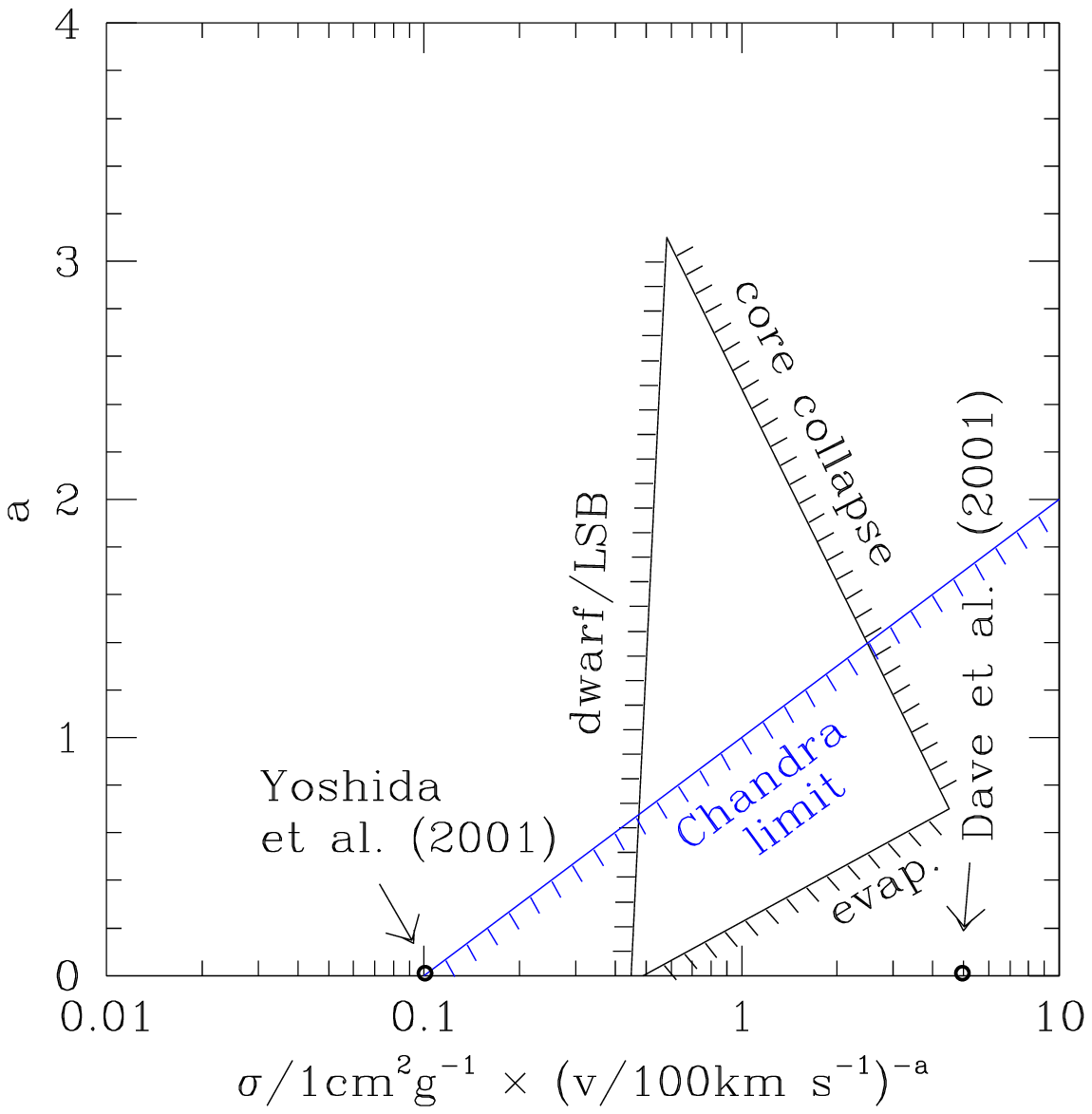}

\figcaption{Constraints on the SIDM cross section, adapted from
\citet{hennawi}.  The hatch marks delineate the region excluded by each
constraint.  The triangular region is derived from several astrophysical
constraints of \citet{hennawi} (see text).  The results of the simulations of
\citet{yoshida} and \citet{dave} are indicated.  The blue excluded region
labeled `Chandra limit' represents the constraint we derive from the absence of
a significant core in the mass profile of EMSS 1358+6245.
\label{f14}}



\clearpage
\begin{deluxetable}{ccccc}
\tablewidth{250pt}
\tablecaption{Source annuli statistics.\label{t01}}
\tablehead{
\colhead{annulus} &
\colhead{R$_{\rm pixel}$} &
\colhead{R$_{\rm kpc}$} &
\colhead{$^{\rm total}_{\rm photons}$} &
\colhead{$^{\rm source}_{\rm photons}$}
}
\startdata
  1 &  11.5  &  32.7  &  2019 & 2004 \\
  2 &  22.9  &  65.2  &  2004 & 1958 \\
  3 &  34.7  &  98.8  &  2005 & 1926 \\
  4 &  47.4  & 135.0  &  2007 & 1895 \\
  5 &  61.6  & 175.4  &  2000 & 1820 \\
  6 &  76.5  & 217.8  &  2000 & 1761 \\
  7 &  93.0  & 264.8  &  2007 & 1682 \\
  8 & 110.8  & 315.5  &  2011 & 1589 \\
  9 & 153.8  & 437.9  &  4013 & 2691 \\
 10 & 285.8  & 813.8  & 10040 & 3299 \\
\enddata
\end{deluxetable}

\clearpage
\begin{center}
\begin{deluxetable}{ccll}
\tablewidth{310pt}
\tablecaption{Comparison of hot and cold gas in the inner two
shells.\label{t02}}
\tablehead{
\colhead{quantity} &
\colhead{component} &
\colhead{shell 1} &
\colhead{shell 2}
}
\startdata
$T/{\rm keV}$ &  cold  & $1.83_{-0.13}^{+0.18}$  & $1.26_{-0.13}^{+0.18}$ \\
              &   hot  & $7.37_{-2.07}^{+5.40}$  & $7.04_{-1.54}^{+2.26}$ \\
& & \\
$\rho/10^{-26}$ g cm$^{-3}$  &  cold  &
  $19.4\pz_{-9.7}^{+0.5}$         &  $14.4\pz_{-7.2}^{+1.5}$   \\
                             &   hot  &
  ${\pz}4.83_{-2.42}^{+0.24}$  &  ${\pz}2.57_{-0.06}^{+0.08}$ \\
& & \\
$\delta M/10^{11}$ \Msun  &  cold  &
  $0.513_{-0.257}^{+0.013}$   &  $0.264_{-0.132}^{+0.027}$  \\
                          &  hot   &
  $0.923_{-0.462}^{+0.047}$   &  $3.81\pz_{-0.076}^{+0.105}$  \\
& & \\
$V/V_{\rm total}$ & cold & 0.121 & 0.012  \\
                  & hot  & 0.879 & 0.988  \\
& & \\
\enddata
\end{deluxetable}
\end{center}

\end{document}